\documentclass[aps,prb,twocolumn,superscriptaddress]{revtex4-1}

\usepackage{amsfonts}	
\usepackage{amsmath}    
\usepackage{bm}
\usepackage{graphicx}   
\usepackage{verbatim}   
\usepackage{color}      
\usepackage[subrefformat=parens,labelformat=parens,caption=false]{subfig}	
\usepackage{hyperref}   
\usepackage{mathtools}

\usepackage{todonotes}

\newcommand{\lef}{\left(}
\newcommand{\rig}{\right)}

\newcommand{\imu}{\mathrm i}

\newcommand{\deriv}{\mathrm d}

\newcommand{\e}{\mathrm e}
\newcommand{\kB}{{k_\mathrm{B}}}

\newcommand{\EF}{{E_\mathrm{F}}}
\newcommand{\vF}{\upsilon_\mathrm{F}}

\newcommand{\vecr}{\mathbf r}

\newcommand{\veck}{\mathbf k}
\newcommand{\vecv}{\bm{\upsilon}}
\newcommand{\vecK}{\mathbf K}
\newcommand{\vecq}{\mathbf q}
\newcommand{\vecG}{\mathbf G}

\newcommand{\SER}{\mathrm{SER}}
\newcommand{\VAER}{V_\mathrm{AER}}
\newcommand{\AER}{\mathrm{AER}}

\newcommand{\GNR}{{\mathrm{GNR}}}

\newcommand{\pseudo}{{(\mathrm{pp})}}

\newcommand{\diff}{{\mathrm{DE}}}

\newcommand{\angleRibbon}{\gamma_\mathrm{GNR}}

\newcommand{\atomC}{{(\mathrm{C})}}
\newcommand{\atomH}{{(\mathrm{H})}}

\DeclareMathOperator{\arctanh}{arctanh}

\newcommand{\SP}{P}
\newcommand{\SERSD}{\sigma_\SER}
\newcommand{\SERCL}{\Lambda_\SER}
\newcommand{\SERMFP}{\lambda_\SER}
\newcommand{\AERSD}{\sigma_\AER}
\newcommand{\AERCL}{{\Lambda_\AER}}

\begin{document}

\title{Theoretical study of scattering in graphene ribbons \\ in the presence of structural and atomistic edge roughness}
\author{Kristof Moors}
\email[E-mail: ]{kristof.moors@uni.lu}
\affiliation{Physics and Materials Science Research Unit, University of Luxembourg, Avenue de la Fa\"iencerie 162a, L-1511 Luxembourg, Luxembourg}
\author{Antonino Contino}
\affiliation{Electrical Engineering (ESAT) Department, KU Leuven, Kasteelpark Arenberg 10, B-3001 Leuven, Belgium}
\affiliation{imec, Kapeldreef 75, B-3001 Leuven, Belgium}
\author{Maarten L.\ Van de Put}
\affiliation{Department of Materials Science and Engineering, The University of Texas at Dallas, Richardson, Texas 75080, USA}
\author{William G.\ Vandenberghe}
\affiliation{Department of Materials Science and Engineering, The University of Texas at Dallas, Richardson, Texas 75080, USA}
\author{Massimo V.\ Fischetti}
\affiliation{Department of Materials Science and Engineering, The University of Texas at Dallas, Richardson, Texas 75080, USA}
\author{Wim Magnus}
\affiliation{imec, Kapeldreef 75, B-3001 Leuven, Belgium}
\affiliation{Physics Department, University of Antwerp, Groenenborgerlaan 171, B-2020 Antwerpen, Belgium}
\author{Bart Sor\'ee}
\affiliation{imec, Kapeldreef 75, B-3001 Leuven, Belgium}
\affiliation{Physics Department, University of Antwerp, Groenenborgerlaan 171, B-2020 Antwerpen, Belgium}
\affiliation{Electrical Engineering (ESAT) Department, KU Leuven, Kasteelpark Arenberg 10, B-3001 Leuven, Belgium}

\date{\today}

\begin{abstract}
	We investigate the diffusive electron-transport properties of charge-doped graphene ribbons and nanoribbons with imperfect edges.
	We consider different regimes of edge scattering, ranging from wide graphene ribbons with (partially) diffusive edge scattering to ribbons with large width variations and nanoribbons with atomistic edge roughness.
	For the latter, we introduce an approach based on pseudopotentials, allowing for an atomistic treatment of the band structure and the scattering potential, on the self-consistent solution of the Boltzmann transport equation within the relaxation-time approximation and taking into account the edge-roughness properties and statistics.
	The resulting resistivity depends strongly on the ribbon orientation, with zigzag (armchair) ribbons showing the smallest (largest) resistivity, and intermediate ribbon orientations exhibiting intermediate resistivity values.
	The results also show clear resistivity peaks, corresponding to peaks in the density of states due to the confinement-induced subband quantization, except for armchair-edge ribbons that show a very strong width dependence because of their \textit{claromatic} behavior.
	Furthermore, we identify a strong interplay between the relative position of the two valleys of graphene along the transport direction, the correlation profile of the atomistic edge roughness, and the chiral valley modes, leading to a peculiar strongly suppressed resistivity regime, most pronounced for the zigzag orientation.
\end{abstract}

\maketitle

\section{Introduction} \label{sec:introduction}
Fujita \textit{et al.}~first introduced graphene ribbons in 1996\cite{Fujita1996, Nakada1996, Wakabayashi1999} and, at present, they can be produced using a variety of techniques, e.g., lithography,\cite{Chen2007, Han2007} chemical processing,\cite{Datta2008, Campos-Delgado2008, Yang2008, Li2008} unzipping or etching of carbon nanotubes,\cite{Jiao2009, Kosynkin2009, Elias2010, Jiao2010} molecular precursors,\cite{Cai2010} ion implantation,\cite{Tongay2012} or exfoliation.\cite{Mohanty2012}
The intrinsically very promising properties of two-dimensional graphene (high mechanical stability and large thermal and electrical conductivity) have led many to propose alternative device and integrated circuit components in which charge-doped graphene nanoribbons are employed in order to meet future very-large-scale integration standards. A notable application is nanoscale interconnects, for which a low electrical resistivity and high current density is crucial.\cite{Naeemi2007, Murali2009, Li2009, Chuan2009, Rakheja2013}
The degradation of the mobility of graphene ribbons for decreasing ribbon widths, due to diffusive edge scattering processes, forms a crucial issue in that regard, hampering the large-scale integration of nanoscaled graphene ribbon devices.\cite{Murali2009}

Several publications have already addressed the (diffusive) transport properties of graphene ribbons for different scattering mechanisms, e.g., impurity, edge (roughness or disorder) or acoustic and optical phonon scattering.\cite{Naeemi2007, Cresti2008, Fang2008, Chuan2009, Li2009, Cresti2009, Bresciani2010, Takane2010, Fischetti2011, Xu2012, Fischetti2013, Dugaev2013, Rakheja2013, Misawa2015} The approaches that were considered vary widely, ranging from a nearest-neighbor tight-binding description for the band structure of a graphene ribbon near the charge-neutrality (Dirac) point to full-fledged atomistic simulations, e.g., based on empirical pseudopotentials.\cite{Fischetti2013} Similarly, the treatment for these scattering mechanisms ranges from phenomenological or (semi)classical estimates of the mean free path, entering the Landauer conductance formula, to numerically solving Green's functions and perturbative or atomistic scattering approaches in combination with the Boltzmann transport equation.

In this work, we revisit and extend the existing treatments of edge scattering, which is typically the dominant scattering mechanism for charge-doped graphene ribbons with imperfect edges and becomes increasingly important as the ribbon width decreases.\cite{Murali2009} Our main aim is to obtain a general description for edge-roughness scattering in graphene nanoribbons with arbitrary orientation (zigzag, armchair, or otherwise) and study the impact of different parameters, including the edge profile properties and statistics, the doping level, and the ribbon width (ranging from the $\mu$m scale down to the nm scale) and orientation, on their transport properties.

In the case of wide graphene ribbons, a continuum description for the in-plane momenta of the electrons is satisfactory. If, furthermore, the width variations are relatively small, edge scattering can be described by a phenomenological probability for diffusive edge scattering, as initially proposed by Naeemi \textit{et al}.\cite{Naeemi2007} This approach can also be extended to account for large width variations, as was done recently by Contino \textit{et al}.\cite{Contino2018}

The major part of this work presents the application of an empirical pseudopotential approach, based on earlier work by Fischetti \textit{et al.} for armchair ribbons with atomistic line-edge roughness,\cite{Fischetti2011} to graphene nanoribbons with any orientation. For our purposes, this approach strikes the perfect balance between accuracy and computational burden (compared, for example, to nearest-neighbor tight-binding models and self-consistent pseudopotential approaches based on density functional theory). A crucial novelty of the approach presented here is the consideration of the Boltzmann transport equation within the self-consistent relaxation-time approximation. This formalism has frequently been simplified in previous treatments, whereas we show that this can lead to large errors of the resulting resistivity.

We also propose a simplified description for atomistic edge-roughness scattering, based on the model by Brey \textit{et al}.,\cite{Brey2006B} and on the edge-roughness statistics, to greatly reduce the computational burden with respect to the pseudopotential approach. The simplified model can be fitted to excellent quantitative agreement with the pseudopotential approach and allows for a systematic comparison with the other scattering models for different ribbon orientations and edge profile parameters over a wide range of ribbon widths.

The paper is structured as follows. In Sec.~\ref{sec:model}, we introduce the different descriptions of the electric charge carriers in wide and narrow graphene ribbons. In Sec.~\ref{sec:scattering}, we cover the different edge-scattering models, ranging from diffusive edge scattering in wide ribbons to atomistic edge-roughness scattering. Section~\ref{sec:transport} deals with the transport formalism for the different scattering mechanisms. A comparison and discussion of atomistic edge roughness, structural edge roughness, and diffusive-edge scattering for the different types of ribbons is presented in Sec.~\ref{sec:comparison}, before concluding and providing a brief outlook in Sec.~\ref{sec:conclusion}.

\section{Graphene ribbons} \label{sec:model}
In this section, we present the two different approaches to model the electronic structure of graphene (nano)ribbons.
Our main goal is to obtain the appropriate dispersion relation $E(\veck)$ for the (nano)ribbon under consideration, from which the wave vector, group velocity, and density of states of the different Fermi level states can be extracted in a straightforward manner and used for transport modeling.

\begin{figure*}[htb]
	\centering
	\subfloat[\ \label{fig:graphene_a}]{\includegraphics[height=0.19\linewidth]{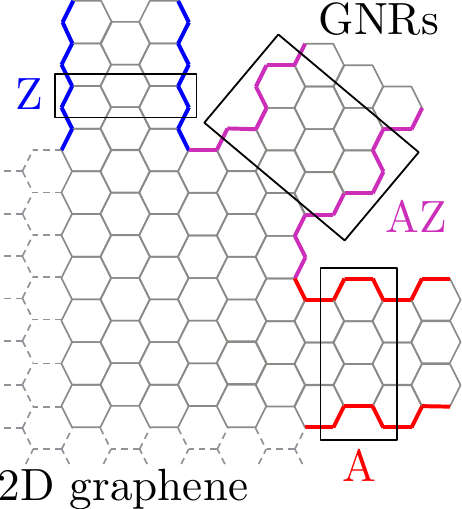}}
	\hspace{0.01\linewidth}
	\subfloat[\ \label{fig:graphene_b}]{\includegraphics[height=0.19\linewidth]{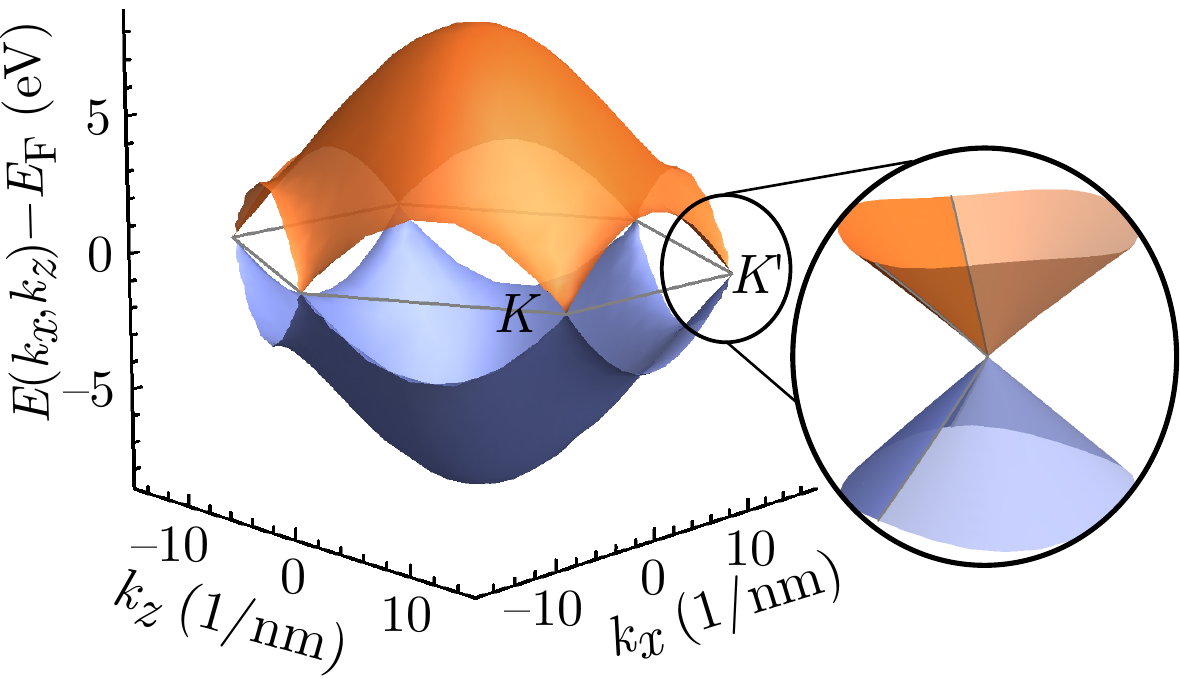}}
	\hspace{0.01\linewidth}
	\subfloat[\ \label{fig:graphene_c}]{\includegraphics[height=0.19\linewidth]{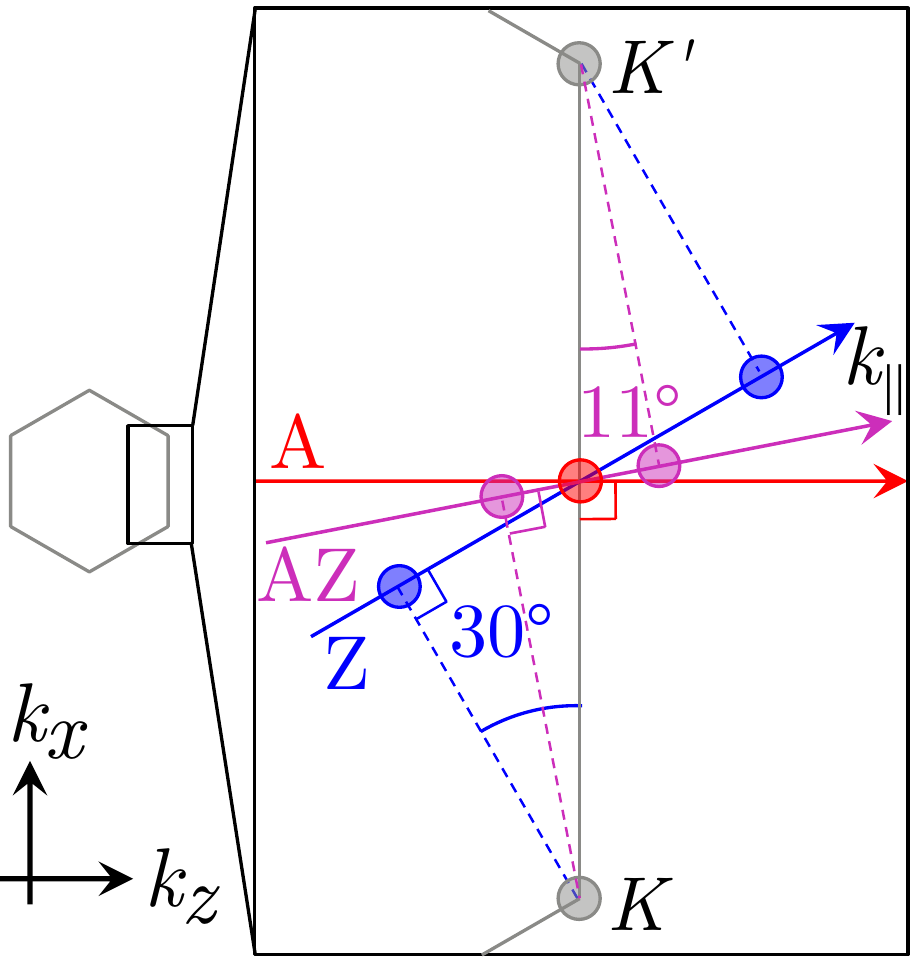}}
	\hspace{0.01\linewidth}
	\subfloat[\ \label{fig:graphene_d}]{\includegraphics[height=0.15\linewidth]{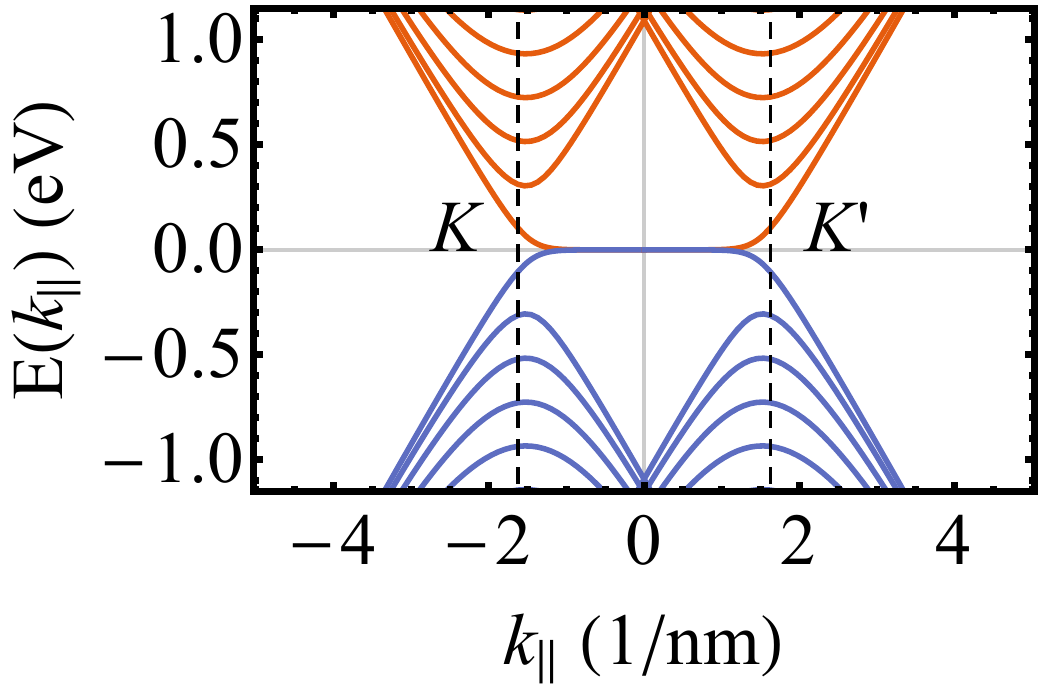}}
	\caption{
		(a) A schematic representation of three possible GNR orientations with supercell indicated by a rectangle: zigzag (Z), armchair (A), and armchair-zigzag (AZ).
		(b) The $\pi$-bond bands of graphene are shown as a function of the wave numbers $k_x$ and $k_z$ in the hexagonal first Brillouin zone, as obtained from a nearest-neighbor tight binding model, leading to Dirac cones at special points $K$ and $K^\prime$.
		(c) A visualization of the Dirac cone projection procedure, explained in Appendix~\ref{subsec:GNR_simplified}, for the ribbon orientations depicted in (a).
		(d) The energy spectrum for a $\sim$10-nm-wide armchair-zigzag GNR, consisting of bulk conduction (orange) and valence (blue) subbands, as well as edge states (black), according to the simplified GNR model of Appendix~\ref{subsec:GNR_simplified}.
	}
	\label{fig:graphene}
\end{figure*}

\subsection{Wide ribbons} \label{subsec:wide_ribbon}
When a graphene ribbon is sufficiently wide (compared to the electron phase-coherence length), its electronic band structure can be described by that of bulk graphene for transport-modeling purposes.
Indeed, for ribbon widths down to a few tens of nanometer, the spectrum will typically consist of narrowly spaced subbands, which one can approximate by a continuum of wave vectors. Figure~\subref*{fig:graphene_a} shows the classic honeycomb lattice of graphene, together with its well known band structure [see Fig.~\subref*{fig:graphene_b}]. Near the Dirac point energy, the two-dimensional (2D) graphene dispersion relation is linear and equal to
\begin{align} \label{eq:2D_disp}
	E_\vecK(\veck) = \pm \hbar \vF |\veck - \vecK|, \quad
		E_{\vecK^\prime}(\veck) = \pm \hbar \vF |\veck - \vecK^\prime|,
\end{align}
where $\vF$ is the Fermi velocity in graphene, approximately equal to $1\times10^6$~m/s and $\vecK$, $\vecK'$ are two high-symmetry points at the edge of the Brillouin zone where the valence and conduction bands touch [see Fig.~\subref*{fig:graphene_a}]: $\vecK \equiv (K_x, K_z) = (-\sqrt{3}/3, 1) 2 \pi / (3 a_0)$, $\vecK' = (\sqrt{3}/3,1) 2 \pi / (3 a_0)$, with $a_0 \approx 0.142$~nm the bond length between two neighboring carbon atoms. The 2D density of states can be evaluated analytically and is equal to
\begin{align} \label{eq:2D_DOS}
	\frac{\deriv n}{\deriv E} = g_s g_v \frac{E}{2 \pi \hbar^2 \vF^2},
\end{align}
with $g_s = 2$ the spin degeneracy and $g_v =2 $ the valley degeneracy accounting for the two (spin-degenerate) Dirac cones at $K$ and $K'$.

\subsection{Nanoribbons} \label{subsec:GNR}
To study the electronic structure and the (electron) transport properties of arbitrarily oriented [three examples are illustrated in Fig.~\subref*{fig:graphene_a}] graphene nanoribbons (GNRs), we will employ the empirical pseudopotential method that has been successfully applied for various carbon-based structures using pseudopotential parameters introduced by Kurokawa \textit{et al}.\cite{Fischetti2013, Kurokawa2000} With this method, the following system of equations needs to be solved in order to obtain the Bloch wave functions $\psi_{\veck \, n} (\vecr)$:
\begin{align} \label{eq:pseudoGNR} \begin{split}
	&\sum_{\vecG'} \lef \frac{\hbar^2}{2 m_\e}
		\left| \veck + \vecG\right|^2 \delta_{\vecG
		, \vecG'} + V^\pseudo_{\vecG - \vecG'} \rig
		u_{\veck + \vecG' , n} \\
	&\qquad = E_n (\veck) \, u_{\veck + \vecG , n},
	\end{split}
\end{align}
with
\begin{align}
	\begin{split}
		\psi_{\veck , n} (\vecr) &=
			\e^{\imu \veck \cdot \vecr} \,
			u_{\veck , n}(\vecr), \\
		u_{\veck , n}(\vecr) &\equiv \sum_\vecG
			u_{\veck + \vecG , n} \,
				\e^{\imu \vecG \cdot \vecr}.
	\end{split}
\end{align}
We expand the part of the Bloch wave function that is periodic over the supercell, $u_{\veck , n}(\vecr)$, on a plane-wave basis with Fourier components $u_{\veck + \vecG , n}$. $\vecG$ are reciprocal lattice vectors and $V^\pseudo_\vecG$ are the empirical pseudopotential Fourier components.
The integer index $n$ and the wave vector $\veck$ label the different solutions of Eq.~\eqref{eq:pseudoGNR} and time-reversal symmetry and the absence of spin-orbit coupling (being a reasonable assumption for graphene\cite{Gmitra2009}) imposes the following relations for states with equal $n$ and opposite wave vector:\cite{Martin2004}
\begin{align}
\begin{split}
	&\psi_{-\veck , n}(\vecr) = \psi^*_{\veck , n}(\vecr),
		\qquad \quad u_{-\veck + \vecG , n} = u^*_{\veck - \vecG , n}, \\
	&E_n(-\veck) = E_n(\veck).
\end{split}
\end{align}

\begin{figure}[tb]
	\centering
	\subfloat[\ \label{fig:pseudoPotentials_a}]{\includegraphics[width=0.48\linewidth]{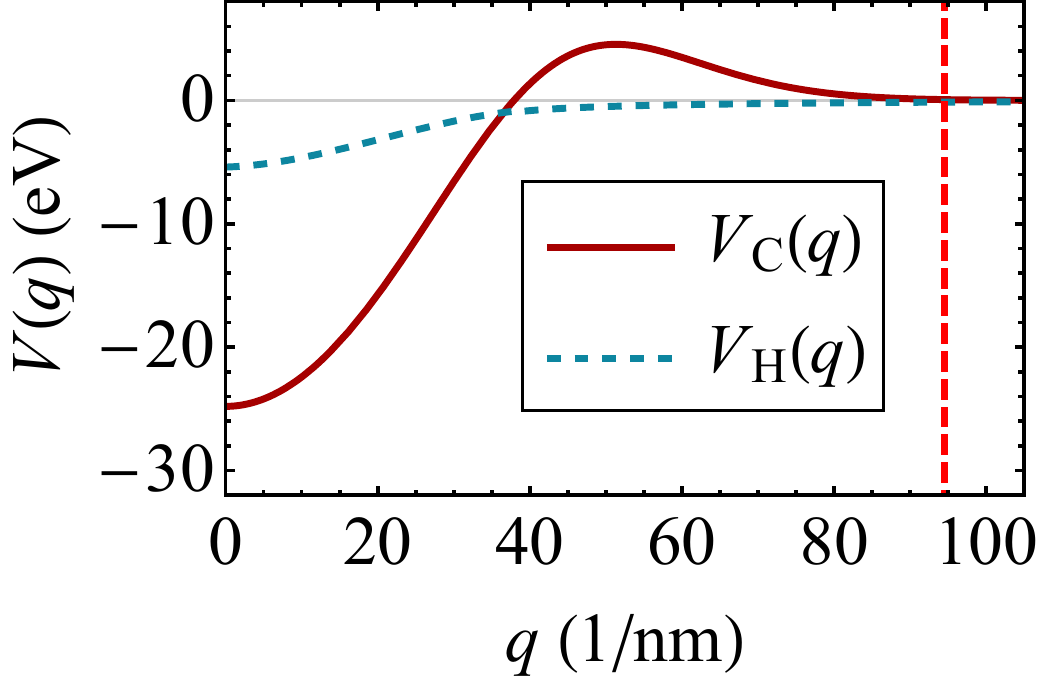}}
	\hspace{0.01\linewidth}
	\subfloat[\ \label{fig:pseudoPotentials_b}]{\includegraphics[width=0.48\linewidth]{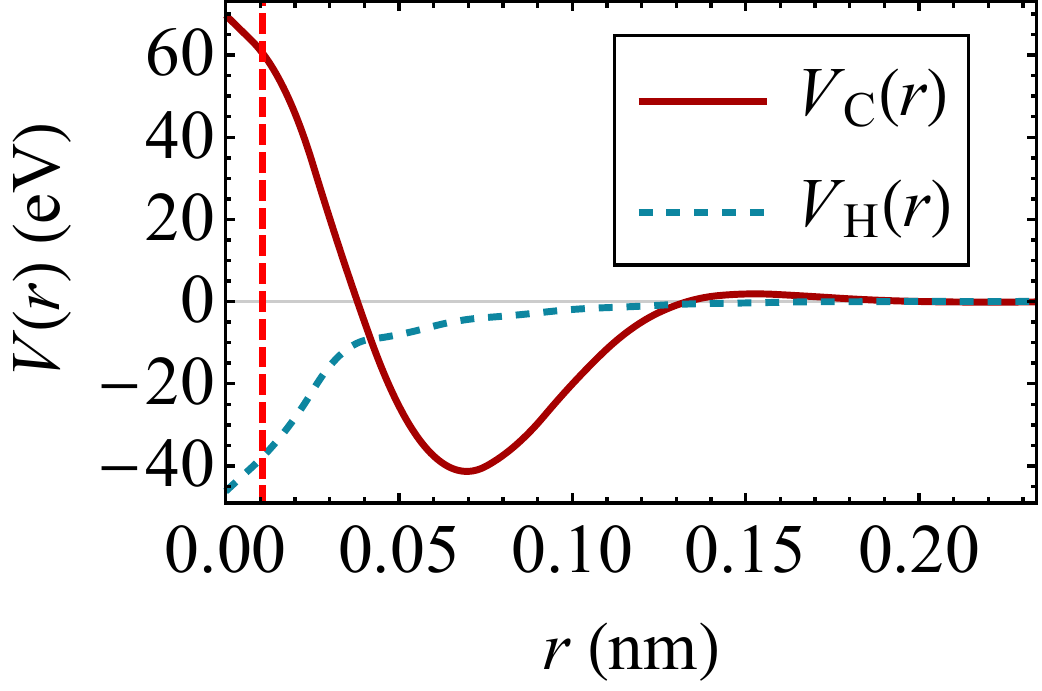}}
	\caption{
	The Kurokawa pseudopotentials of a single carbon and hydrogen atom as defined in Eq.~\eqref{eq:pseudo} are shown in (a) reciprocal (normalized to the atomic volume of C in diamond, approximately equal to 5.7~\AA${}^3$) and (b) real space, using the Kurokawa parameters provided in Table~\ref{table:Kurokawa}. The wave vector cutoff (discretization) is indicated in reciprocal (real) space by the red dashed line.
	}
	\label{fig:pseudoPotentials}
\end{figure}

Our study will be restricted to GNRs with a standard edge termination using hydrogen atoms.\cite{Fischetti2013} We consider the pseudopotentials of single carbon (C) and hydrogen (H) atoms to be isotropic and given by:
\begin{align} \label{eq:pseudo}
\begin{split}
	V^\atomC_q &\equiv
		\frac{b_1 (b_3 q^2 - b_2)}{\exp(b_3 q^2 - b_4) + 1}, \\
	V^\atomH_q &\equiv
		\left\{ \begin{matrix}
			\sum\limits_{i = 1}^3 b_i q^i \qquad (q \leq 2) \\
			\sum\limits_{i = 1}^4 b_{-i} q^{-i} \quad \! \! (q > 2)
		\end{matrix} \right. \! ,
\end{split}
\end{align}
as prescribed by Kurokawa \textit{et al}.~in Rydberg atomic units with parameters listed in Table~\ref{table:Kurokawa}. We present the pseudopotentials in reciprocal and real space in Fig.~\ref{fig:pseudoPotentials}.

\begin{table}[tb]
	\centering
	\caption{Local pseudopotential parameters for C and H from Kurokawa \textit{et al}.~(normalized to the atomic volume of C in diamond, approximately 5.7~\AA${}^3$) in Rydberg atomic units.}
	\begin{tabular}{lccccc}
		\hline
		\hline
 		& \, $b_0$ \, & \, $b_1$ \, & \, $b_2$ \,
 		& \, $b_3$ \, & \, $b_4$ \\
		\hline
		C \, & \, \, & \, 1.781 \, & \, 1.424 \, & \, 0.354 \,
			& \, 0.938 \\
		H \, & \, -0.397 \, & \, 0.02759 \, & \, 0.1754 \,
			& \, -0.0531 \, & \, \\
		\hline
		\hline
		& \, $b_{-1}$ \, & \, $b_{-2}$ \, & \, $b_{-3}$ \, &
		\, $b_{-4}$ \, & \, \\
		\hline
		H \, & 0.0811 \, & \, -1.086 \, & \, 2.71 \,
		& \, -2.86 \, & \, \\
		\hline
		\hline
	\end{tabular}
	\label{table:Kurokawa}
\end{table}

We employ a highly optimized fast Fourier transform and eigenvalue solver developed by Van de Put \textit{et al}.\cite{Frigo2005, Kresse1996, VandePut2016, Wood1985} to obtain numerical solutions of Eq.~\eqref{eq:pseudoGNR} in a discretized $\veck$ space in the positive half of the first Brillouin zone, with an energy cutoff of 25~Ry.
Along the (in-plane and out-of-plane) confinement directions, the supercell contains $\sim$2~nm of vacuum on each side, ensuring that the wave functions are properly contained within the supercell.
The wave functions obtained by solving Eq.~\eqref{eq:pseudoGNR} take the following form:
\begin{align}
	\psi_{\veck_\parallel , n}(\vecr) \equiv
		\e^{\imu \veck_\parallel \cdot \vecr_\parallel} \mkern-35mu \sum_{\vecG \, (|\vecG| \leq G_\mathrm{cutoff})} \mkern-35mu
		u_{\veck_\parallel + \vecG , n} \,
		\e^{\imu \vecG \cdot \vecr},
\end{align}
where $\veck_\parallel$ and $\vecr_\parallel$ are the wave vector and position coordinate along the GNR (transport) direction and $G_\mathrm{cutoff}$ is the wave vector cutoff associated with the energy cutoff (equal to 5 inverse Bohr so that we capture spatial variations down to a length scale of $\sim$0.01~nm). Note that the solutions are independent of the wave vector $\veck_\perp$, perpendicular to the GNR direction, as long as the energy is smaller than the vacuum level.

\begin{figure}[tb]
	\centering
	\subfloat[\ \label{fig:GNR_Bands_a}]{\includegraphics[width=0.48\linewidth]{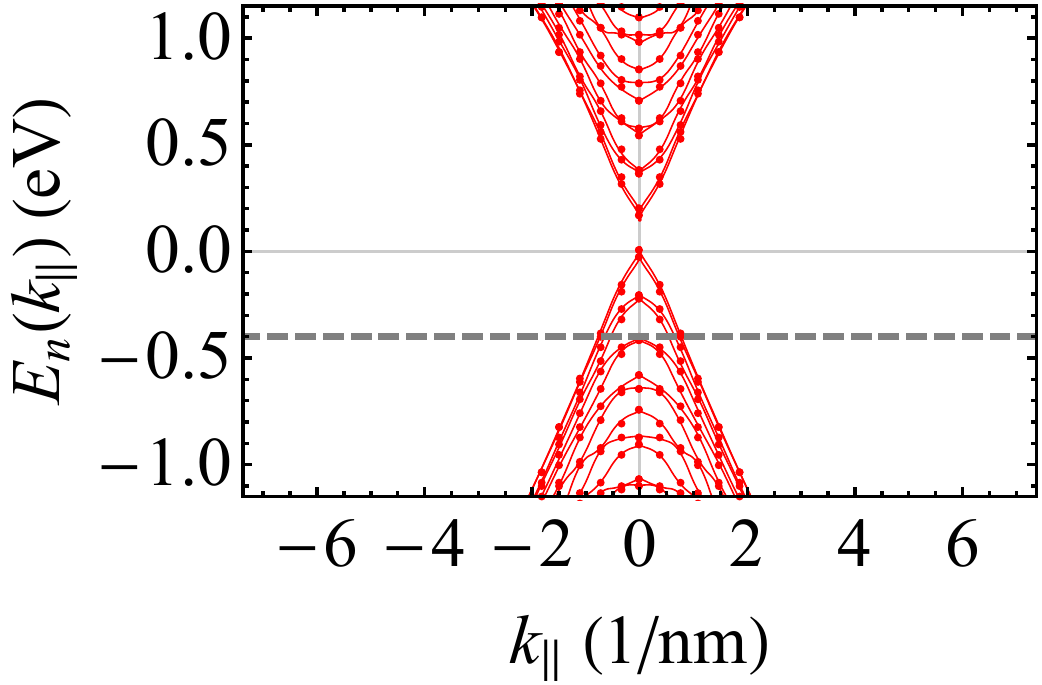}}
	\hspace{0.01\linewidth}
	\subfloat[\ \label{fig:GNR_Bands_b}]{\includegraphics[width=0.48\linewidth]{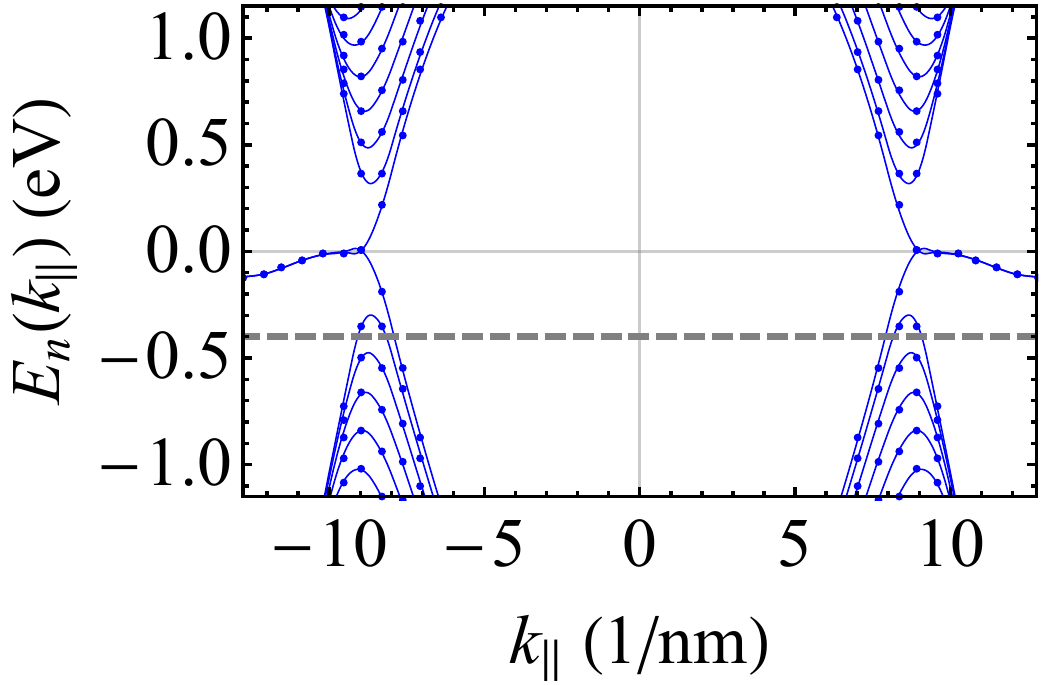}} \hfill
	\subfloat[\ \label{fig:GNR_Bands_c}]{\includegraphics[width=0.48\linewidth]{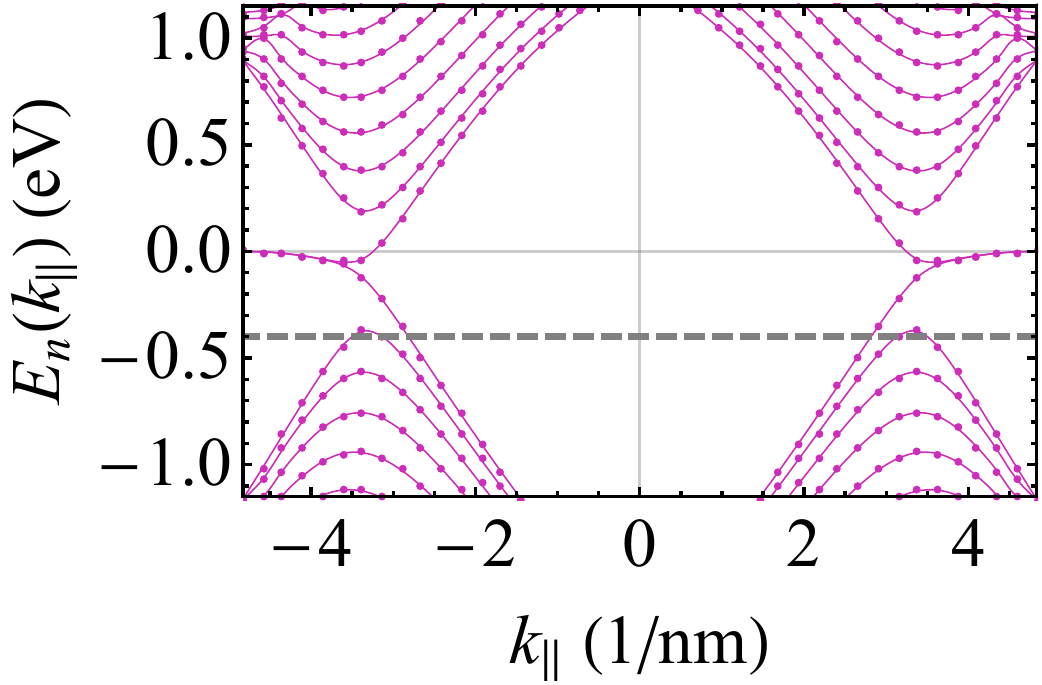}}
	\hspace{0.01\linewidth}
	\subfloat[\ \label{fig:GNR_Bands_d}]{\includegraphics[width=0.48\linewidth]{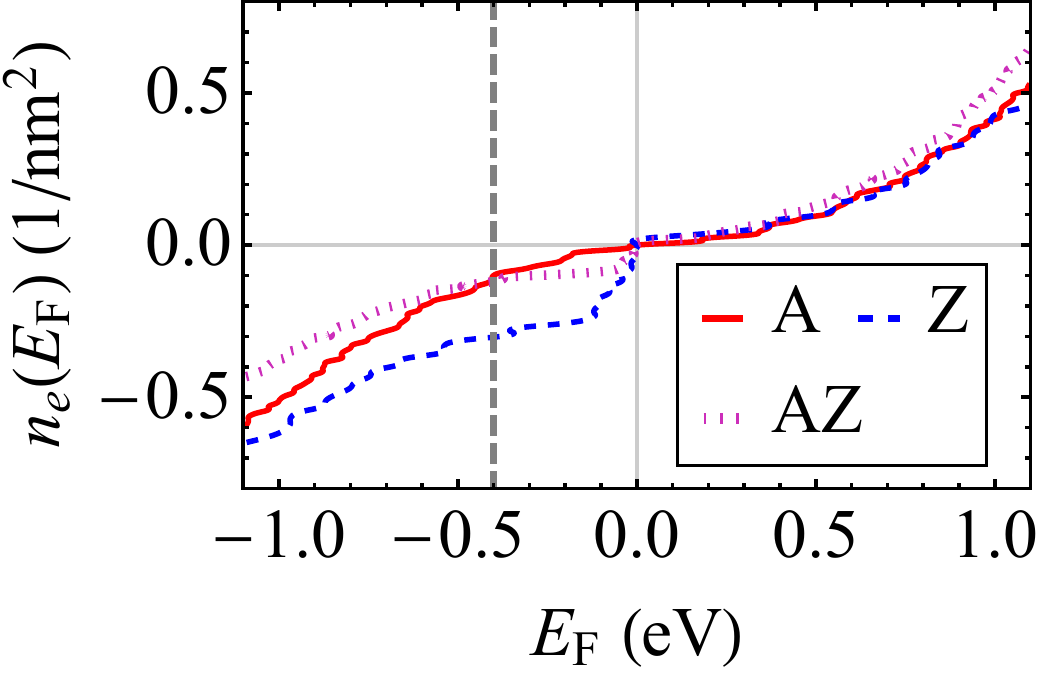}}
	\caption{
		(a)-(c) The (sub)band structures of 10-nm-wide GNRs with (a) armchair (A), (b) zigzag (Z), and (c) armchair-zigzag (AZ) edge configuration [see Fig.~\protect\subref*{fig:graphene_a}], which are obtained with the pseudopotential approach of Sec.~\ref{subsec:GNR}, are presented as a function of the transport wave vector $k_\parallel$.
		(d) The charge density is shown as a function of the Fermi level.
		A gray dashed line intersects at a doping level of $-0.4$~eV with respect to the charge-neutrality point, in all subfigures.
	}
	\label{fig:GNR_Bands}
\end{figure}

In Fig.~\ref{fig:GNR_Bands}, the band structure for three 10-nm-wide GNRs with different orientations is presented.
In addition to the typical armchair and zigzag configurations, we consider the simplest edge variation by alternating armchair and zigzag supercell units [see Fig.~\subref*{fig:graphene_a}]. A GNR with this edge configuration is denoted as an armchair-zigzag GNR and makes an angle of $\sim$11${}^\circ$ with an armchair GNR, while the fraction of zigzag edges is about 36\% along the length of the ribbon.

The armchair GNR has a band structure that closely resembles a one-dimensional projection of a Dirac cone, apart from a small band gap that appears where the Dirac point is expected. This is a confinement-induced gap that, in addition to the typical $1/W^2$ scaling, is very sensitive to the GNR width $W$, due to the spatial distribution of Clar resonance structures.\cite{Clar1972, Balaban2009, Fischetti2013} There are three distinct types of spatial distributions that can be realized in armchair GNRs, and they are cycled through periodically by each number of atomic layers $n_\mathrm{a}$ along the width of the GNR ($n_\mathrm{a} \mod 3$), leading to three distinct confinement-induced band gaps. This effect is also known as the \textit{claromatic} behavior of armchair GNRs.\cite{Fischetti2013}

The zigzag GNR shows a different deviation from the 2D graphene Dirac cones with the appearance of close to dispersionless subbands near the Dirac point. These subbands are well known to arise from localized electron states at the zigzag edges.\cite{Fujita1996} The armchair-zigzag GNR shows similar dispersionless subbands connecting the two separated Dirac cones. As long as the GNR deviates from the armchair orientation, two Dirac cones, which are separated in $k_\parallel$, appear together with edge states.\cite{Ryu2002,Brey2006A,Brey2006B,Akhmerov2008,Delplace2011,VanMiert2016}
The claromatic behavior observed in armchair GNRs does not play a significant role for other GNR orientations. Therefore, the energy spectrum is less sensitive to small width variations, something that we confirmed with additional pseudopotential simulations.
Further note that for graphene nanostructures, it was shown that ferromagnetic ordering can occur at zigzag edges and that an antiferromagnetic interedge exchange coupling can induce a band gap.\cite{Bhowmick2008, Jung2009, Magda2014} Our pseudopotential approach does not capture these effects, but they can safely be neglected for the GNR widths and doping (Fermi) levels (with, correspondingly, substantial charge screening) under consideration.

Apart from the band structure, the total charge density (obtained from the electron density of the different subbands) is presented as a function of the Fermi level $\EF$ in Fig.~\subref*{fig:GNR_Bands_d}, with $\EF = 0$ considered at the charge-neutrality point (i.e., the top of the highest valence subband in case of a gapped band structure). It clearly shows that a Fermi level shift of a few hundred meV requires relatively high charge doping levels, albeit lower than what would be obtained from considering the 2D graphene density of states. Another aspect that can be observed is that a much larger charge density is needed to pull down the Fermi level of the zigzag GNR due to the extended dispersionless band near the charge-neutrality point. The charge density that these edge state subbands host is proportional to the fraction of zigzag-versus-armchair units in the edge configuration.

A nonatomistic approach that is able to reproduce these band structures qualitatively for arbitrary ribbon orientations, based on the Dirac equation with appropriate boundary conditions, is presented in Appendix~\ref{subsec:GNR_simplified}.

\section{Edge scattering}
\label{sec:scattering}
\begin{figure}[tb]
	\centering
	\includegraphics[width=0.98\linewidth]{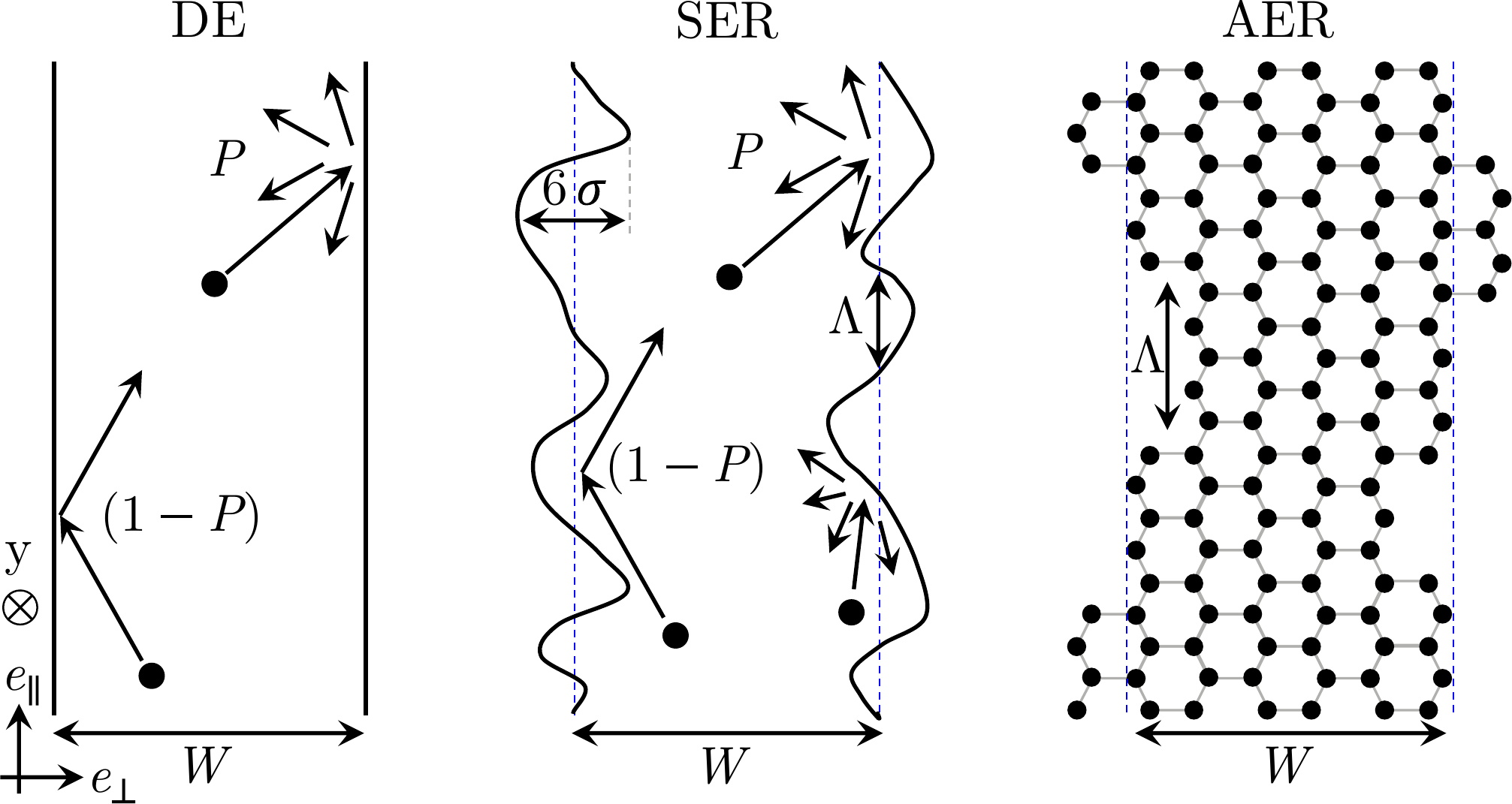}
	\caption{
		The three different types of edge scattering for a graphene (nano)ribbon with (average) width $W$ are represented schematically: DE scattering (left) with edge scattering parameter $\SP$; SER scattering (center) with, in addition, the SER standard deviation $\SERSD$ and correlation length $\SERCL$; AER scattering (right), which is characterized by the AER correlation length $\AERCL$ and the type of edge modifications (only the removal or addition of a single row of hydrogen-terminated carbon atoms here).
	}
	\label{fig:edge_scattering}
\end{figure}

We divide edge scattering into three categories, as depicted in Fig.~\ref{fig:edge_scattering}: diffusive edge (DE), structural edge-roughness (SER), and atomistic edge-roughness (AER) scattering. We treat them separately in Secs.~\ref{subsec:diffusive}, \ref{subsec:SER}, and \ref{subsec:AER}, respectively.

\subsection{Diffusive edge} \label{subsec:diffusive}
We start with the most simple case: The electrons are simply moving forward as well as back and forth between the two edges, with a certain probability of being diffusely scattered [see Fig.~\ref{fig:edge_scattering} (left)]. This is similar to what is presented in other publications,\cite{Chuan2009,Naeemi2007,Li2009,Rakheja2013,Bresciani2010} except that we will just consider the 2D density of states of graphene here and not the quantized 1D density of states of GNRs.

An electron in graphene, traveling with velocity $\vecv$ immediately after impact with an edge, will travel a distance $L_\parallel$ along the transport direction and a distance $L_\perp$ along the direction perpendicular to transport before there is another collision with an edge. $L_\perp$ can be considered equal to the graphene width $W$, such that we obtain:
\begin{align}
	L_\parallel =  L_\perp \frac{|\vecv|\sin\theta}{|\vecv|\cos\theta}  = W \tan\theta,
\end{align}
with $\theta = \arctan(\upsilon_\perp/\upsilon_\parallel)$.
For the collision event at the edge, we consider a probability of $1 - \SP$ for specular scattering, which does not affect the mean free path (MFP), and a probability $\SP$ for diffusive scattering (note that this probability differs from the conventional definition of the specularity parameter in the Fuchs-Sondheimer model for boundary scattering in metallic thin films\cite{Fuchs1938, Sondheimer1952}). At the same time, $L_\parallel/ \SP$ cannot exceed the maximum MFP along the transport direction, given by $\lambda_\mathrm{2D} \sin\theta$, with $\lambda_\mathrm{2D}$ the MFP associated with other scattering mechanisms for bulk graphene such as phonon or impurity scattering. With these considerations in mind, we obtain the total MFP for the case of a (partially) diffusive edge $\lambda_\diff$ by averaging over all angles of the velocity:
\begin{align} \label{eq:lambda_diff}
	\begin{split}
	\lambda_\diff &=\frac{2}{\pi} \mkern-5mu \int\limits_0^{\pi/2} \mkern-7mu \deriv\theta \; \lef\frac{\SP}{L_\parallel} + \frac{1}{\lambda_\mathrm{2D} \sin \theta} \rig^{-1} \\
	&= \frac{2W}{\pi \SP} \ln (1 + \SP \lambda_\mathrm{2D} / W).
	\end{split}
\end{align}
Note that the same theory applies also for the case of GNRs, except that, because of the quantization of the density of states, the integral over all possible angles becomes a summation over the available subbands.

\subsection{Structural edge roughness} \label{subsec:SER}
In the case of large edge variations, the mobility degradation due to edge scattering is so large that merely considering a ribbon with uniform width and diffusive scattering at the edge is not enough. In a recent Letter, Contino \textit{et al}.\cite{Contino2018}~presented a model to overcome this problem by including the width variations of the ribbon. Here, we summarize their model and present the corresponding formula for the MFP.

We divide the GNR into two regions: the edge region, where electrons are mostly assumed to impact with the same edge more than once, and the inside region, in the bulk of the ribbon, where electrons can impact with either of the two edges, depending on the electron transport angle [see Fig.~\ref{fig:edge_scattering} (center)]. The distance traveled by electrons in the edge region before impacting with the edge is very random and it strongly depends on the edge profile. However, if we consider only the average distance, we can assume this value to be equal to the autocorrelation length $\SERCL$ of the edge roughness. In the inside region, instead, the same theory used to derive Eq.~\eqref{eq:lambda_diff} can be used, except that $L_\perp$ is not a constant value, equal to the graphene width $W$, but it depends on both the direction of the velocity of the electron and its position in the direction perpendicular to the transport direction $r_\perp$.
By averaging and summing the MFP of both regions, we retrieve the following MFP for SER scattering $\SERMFP$:
\begin{widetext}
\begin{align} \label{eq:mod_final}
	\begin{split}
		\SERMFP &= \frac{1}{\pi \SP (W + 6\SERSD)} \left[-6 P \lambda_\mathrm{2D} \SERSD \vphantom{\frac{P\lambda_\mathrm{2D}+3\SERSD}{W+P\lambda_\mathrm{2D}-3\SERSD}}
			+ 6 \pi P  \SERCL \SERSD + 9 \SERSD^2 \ln \lef \frac{3 \SERSD}{P \lambda_\mathrm{2D} + 3 \SERSD} \rig \right. \\
		&\quad \left. + 2 (W - 3 \SERSD)^2 \arctanh \lef \frac{P \lambda_\mathrm{2D}}{2W + P \lambda_\mathrm{2D} - 6\SERSD} \rig
			+ P W \lambda_\mathrm{2D} + P^2 \lambda_\mathrm{2D}^2 \ln \lef \frac{P \lambda_\mathrm{2D} + 3\SERSD}{W + P \lambda_\mathrm{2D} - 3\SERSD} \rig \right].
	\end{split}
\end{align}
This formula was successfully applied to describe the mobility degradation of graphene ribbon samples ranging from a width of 5~$\mu$m down to 50~nm, with a clear deviation from the DE scattering model for widths below 500~nm due to the large width variations.\cite{Contino2018}

\subsection{Atomistic edge roughness} \label{subsec:AER}
For AER, we will limit ourselves to modifications of the GNR by one row of carbon atoms extra or less on each side [see Fig.~\ref{fig:edge_scattering} (right)], also taking into account a proper adjustment of the hydrogen termination. We can write the AER potential as follows:
\begin{align} \label{eq:AER_potential}
	\VAER(\vecr) = \mkern-15mu
		\sum_{e \in \{\mathrm{L}, \mathrm{R}\}} \sum_{m = \pm} \sigma_{e , m}(\vecr_\parallel) \, V_{e , m} (\vecr), \qquad V_{e , m} (\vecr) = \sum_\vecG V_{e , m , \vecG_\parallel} (\vecr_\perp) \, \e^{\imu \vecG_\parallel \cdot \vecr_\parallel},
\end{align}
with $V_{e , m}(\vecr)$ the potentials of the different GNR modifications, periodically repeated in each supercell, and $V_{e , m , \vecG}(\vecr_\perp)$ the Fourier components of an expansion along the transport direction. The symbol $m$ denotes a modification ($m=+/-$: addition/removal of one row of carbon atoms) at the left ($e =$~L) or right ($e =$~R) edge of the GNR (see examples in Fig.~\ref{fig:potentials}). The function $\sigma_{e , m}(\vecr_\parallel)$ is a stepwise function that is equal to one for the positions (or supercells), along the length of the GNR, where the edge modification $m$ at GNR edge $e$ appears, and equal to zero elsewhere.

We consider a Fourier expansion of the AER potentials along the transport direction, with the directions orthogonal to the transport direction evaluated in real space with coordinate $\vecr_\perp$, as it results in the most tractable form for numerical evaluation. The matrix elements between a certain initial ($i$) and final ($f$) state are given by:
\begin{align} \label{eq:potential_AER}
\begin{split}
	\langle i \mid \VAER \mid f \rangle &= \sum_{e, m}
		\int \mkern-3mu \deriv^3 r \; \e^{- \imu \Delta \veck_\parallel \cdot \vecr_\parallel} \, (u^{(i)})^\ast (\vecr) \, \sigma_{e , m} (\vecr_\parallel) V_{e , m} (\vecr) \, u^{(f)} (\vecr) \\
	&\approx L \sum_{e, m} \sigma_{e , m} (\Delta \veck_\parallel)
		\sum_{\vecG_\parallel, \vecG_\parallel^\prime} \int \mkern-3mu \deriv^2 r_\perp \;
		(u^{(i)}_{ \vecG_\parallel})^\ast(\vecr_\perp) \, V_{e , m , \vecG_\parallel - \vecG_\parallel^\prime}(\vecr_\perp) \, u^{(f)}_{\vecG_\parallel^\prime}(\vecr_\perp),
\end{split}
\end{align}
with $L$ the length of the GNR and $\Delta \veck_\parallel \equiv \veck^{(i)}_ \parallel - \veck^{(f)}_\parallel$.
A disordered edge configuration breaks the periodicity of the total potential and this is captured by the function $\sigma_{e , m}(\vecr_\parallel)$ along the length of the GNR.
We have obtained the last line by considering $\sigma_{e , m}(\vecq_\parallel)$ with $\vecq_\parallel$ only within the first Brillouin zone. The modes outside of the first Brillouin zone are dismissed, as the function, by construction, should not vary over the length of a single supercell. The squared matrix elements are given by:
\begin{align}
\begin{split}
	\left| \langle i \mid \VAER \mid f \rangle \right|^2 &= L^2 \sum_{e, m} \sum_{e^\prime, m^\prime} \sigma_{e , m} (\Delta \veck_\parallel) \, \sigma_{e^\prime , m^\prime} (- \Delta \veck_\parallel)
		\langle i \mid V_{e , m} \mid f \rangle_\perp \langle f \mid V_{e^\prime , m^\prime} \mid f \rangle_\perp, \\
	\textnormal{where } \langle i \mid V_{e , m} \mid f \rangle_\perp &\equiv \sum_{\vecG_\parallel, \vecG_\parallel^\prime} \mkern-3mu \int \mkern-3mu \deriv^2 r_\perp \; (u^{(i)}_{\vecG_\parallel})^\ast(\vecr_\perp)
		\, V_{e , m , \vecG_\parallel - \vecG_\parallel^\prime}(\vecr_\perp) \, u^{(f)}_{\vecG_\parallel^\prime}(\vecr_\perp).
\end{split}
\end{align}

\begin{figure*}[tb]
	\centering
	\raisebox{0.05\linewidth}{\subfloat[\ \label{fig:potentials_a}]{\includegraphics[height=0.13\linewidth]{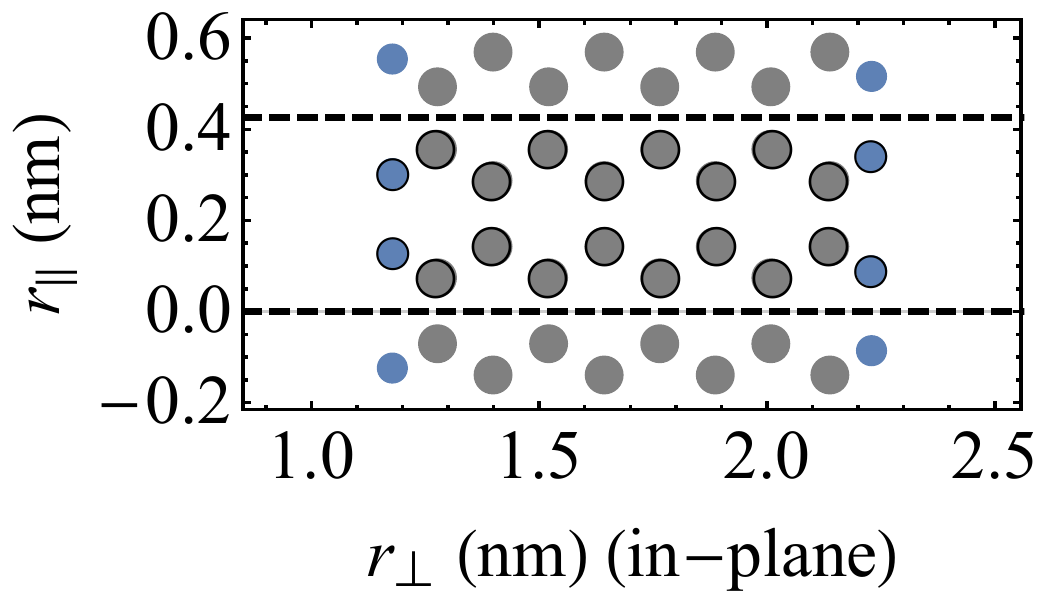}}}
	\hspace{0.0\linewidth}
	\subfloat[\ \label{fig:potentials_b}]{\includegraphics[height=0.235\linewidth]{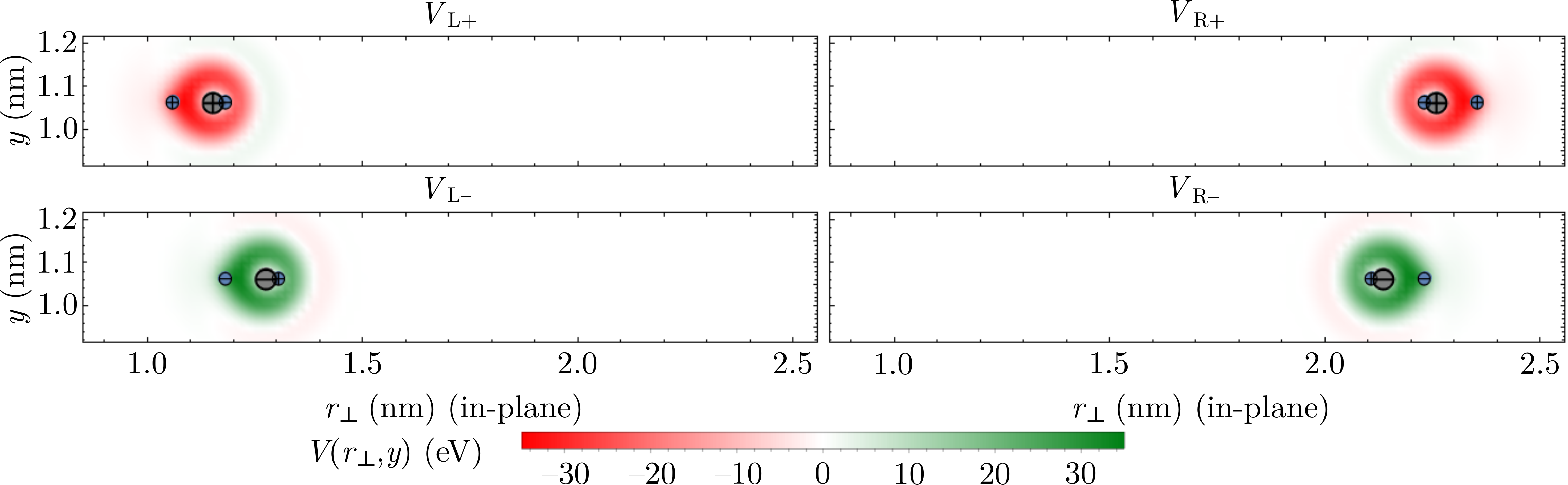}}
	\caption{
		(a) Top view of a $\sim$1-nm-wide armchair GNR (C atoms in gray) with standard hydrogen (blue circles) termination. The supercell is delineated by the black dashed lines.
		(b) The four different AER potentials [see Eq.~\eqref{eq:AER_potential}] of an approximately 1-nm-wide GNR are plotted in the transversal plane, with the depicted values being the average along the transport direction. The atom positions of the edge modifications are superimposed on the figure, the large gray (small blue) circles representing C (H) atoms and $+$ ($-$) denoting an addition (removal) with respect to the unmodified edge.
	}
	\label{fig:potentials}
\end{figure*}

Similar to the description for SER, we model AER by considering its statistical properties, translated to the functions $\sigma_{e , m}(\vecr_\parallel)$. We have only considered the addition or removal of a single row of atoms in each supercell, and we will consider these modifications to be equally probable as no modification (probability of the three cases equal to 1/3), which fixes the average and standard deviation. We consider the two-point correlation function to be Gaussian with a correlation length that should be at least of the order of the length of the supercell for consistency. This also justifies the omission of the higher Fourier modes in the calculation above. The expectation value of the product of roughness functions appearing in the absolute value squared of the matrix elements is accordingly given by:
\begin{align}
	\begin{split}
		\langle \sigma_{e , m}(\vecr_\parallel) \, \sigma_{e^\prime , m^\prime}(\vecr_\parallel^\prime) \rangle_\AER &=
			\left\{ \begin{matrix}
				[ 1 + 2 C_\AERCL(\vecr_\parallel - \vecr_\parallel^\prime)] / 9
					& \textnormal{if } e=e^\prime
						\textnormal{ and } m = m^\prime \\
				[ 1 - C_\AERCL(\vecr_\parallel - \vecr_\parallel^\prime)] / 9
					& \, \, \, \, \textnormal{if } e=e^\prime
						\textnormal{ and } m = - m^\prime \\
				1/9 & \textnormal{otherwise}
			\end{matrix} \right. , \\
		\Rightarrow \langle \sigma_{e , m}(\veck_\parallel) \, \sigma_{e^\prime , m^\prime}(\veck_\parallel^\prime)
			\rangle_\AER &= \left\{ \begin{matrix}
				\delta_{\veck_\parallel + \veck_\parallel^\prime , \mathbf{0}} [ \delta_{\veck_\parallel , \mathbf{0}}
					+ 2 \sqrt{\pi/2} \, \AERCL \,
					C_\AERCL (\veck_\parallel) / (9 L) ] & \textnormal{if } e=e^\prime
						\textnormal{ and } m = m^\prime \\
				\delta_{\veck_\parallel + \veck_\parallel^\prime , \mathbf{0}} [ \delta_{\veck_\parallel , \mathbf{0}}
					- \sqrt{\pi/2} \, \AERCL \,
					C_\AERCL(\veck_\parallel) / (9 L) ] & \, \, \, \, \textnormal{if } e=e^\prime
					\textnormal{ and } m = - m^\prime \\
				\delta_{\veck_\parallel + \veck_\parallel^\prime , 0} \, \delta_{\veck_\parallel , \mathbf{0}} / 9 & \textnormal{otherwise}
			\end{matrix} \right. ,
	\end{split}
\end{align}
\end{widetext}
presented both in real and reciprocal space. We have introduced the Gaussian two-point correlation function, defined as follows:
\begin{align} \label{eq:correlation}
	\begin{split}
		C_\Lambda(\vecr_\parallel - \vecr_\parallel^\prime) &\equiv \exp[  -|\vecr_\parallel - \vecr_\parallel^\prime|^2/(\Lambda^2/2) ], \\
		C_\Lambda(\Delta \veck_\parallel) &\equiv \exp(-|\Delta \veck_\parallel|^2 \Lambda^2 / 8).
	\end{split}
\end{align}
Roughness at the opposite edges is, as before, considered to be fully uncorrelated.
The average of the matrix elements squared, for $\Delta \veck_\parallel \neq \mathbf{0}$ as an example, can thus be expressed as:
\begin{align} \label{eq:ER_Matrix_El}
	\begin{split}
		&\left\langle \left| \langle i \mid \VAER \mid f \rangle \right|^2 \right\rangle_\AER \\
		&\; = \frac{L^2}{9} \mkern-7mu \sum_{e, m, m^\prime} \! \lef \frac{1}{2} + \frac{3}{2} m m^\prime \rig \sqrt{\frac{\pi}{2}} \frac{\AERCL}{L} C_\AERCL(\Delta \veck_\parallel) \\
		&\; \qquad \qquad \times \langle i \mid V_{e , m} \mid f \rangle_\perp \langle f \mid V_{e , m^\prime} \mid f \rangle_\perp.
	\end{split}
\end{align}
Note that a Gaussian roughness profile is only one of many possible choices.\cite{Ando1982} For the Si/SiO${}_2$ interface for example, an exponential power-spectrum is also commonly used.\cite{Goodnick1985, Yu2002} The only change that is required in our model is an adjustment of the two-point correlation function of Eq.~\eqref{eq:correlation} in Eq.~\eqref{eq:ER_Matrix_El}.
Further note that this approach is inspired by the widely used approach to model quantum size effects due to surface roughness in thin metallic films and two-dimensional electron gases, based on the parametrization of boundary deviations of a quantum well and the statistics thereof [here replaced by the functions $\sigma_{e , m}(\vecr_\parallel)$ and their statistics].\cite{Ando1982, Prange1968, Tesanovic1986, Trivedi1988, Meyerovich1994}

A simplified expression for the result in Eq.~\eqref{eq:ER_Matrix_El} that is compatible with the simplified GNR model of Appendix~\ref{subsec:GNR_simplified} is presented in Appendix~\ref{subsec:AER_simplified}.

\section{Transport}
\label{sec:transport}
\textit{A priori}, one can expect the electric charge transport of a significantly charge-doped and sufficiently long, narrow graphene ribbon with imperfect edges to be in the diffusive regime, with diffusive scattering being dominated by edge-scattering events.
The overall impact of edge scattering on the transport behavior is captured by the 2D resistivity $\rho$, which can be obtained from the MFP $\lambda = \lambda_{\diff \, (\SER)}$ due to DE (SER) scattering through $\mu = \lambda e n$, with $\mu = \mu_{\diff \, (\SER)}$ the 2D mobility, $e$ the elementary charge, and $n$ the carrier density. The 2D resistivity is then given by $\rho = 1/\mu$.

In case of AER scattering, the resistivity can be obtained from the Boltzmann transport equation with the self-consistent relaxation-time approximation:\cite{Jacoboni2010, Moors2014}
\begin{align} \label{eq:BTE}
	\frac{1}{\tau_i} = \sum_f \lef 1 - \frac{\upsilon_f}{\upsilon_i} \frac{\tau_f}{\tau_i} \rig \mathcal{P}(i \leftrightarrow f),
\end{align}
with $\upsilon_{i \, (f)}$ the group velocity of the initial (final) state and $\tau_i, \tau_f$ their relaxation times, which can be self-consistently obtained without further approximations. $\mathcal{P}(i \leftrightarrow f)$ is the scattering rate between the two states, obtained from the pseudopotential based [see Eq.~\eqref{eq:ER_Matrix_El}] or simplified [see Eq.~\eqref{eq:AER_scat_simplified}] expression for the matrix element squared through Fermi's golden rule. The simplified model reduces significantly the computational burden, so that the resistivity can easily be evaluated for different ribbon widths and orientations and can be systematically compared to DE and SER scattering. Furthermore, rather than extending the AER potential in Eq.~\eqref{eq:AER_potential} for larger edge modifications, one can consider the typical scaling behavior, $\rho^\prime/\rho \propto (\AERSD^\prime/\AERSD)^2$ [see Eq.~\eqref{eq:AER_scat_simplified}]. One should of course keep in mind that this is a perturbative approach, which is valid only for small-enough width variations.\cite{Moors2016}

\begin{figure*}[htb]
	\centering
	\subfloat[\ \label{fig:GNR_ER_a}]{\includegraphics[height=0.145\linewidth]{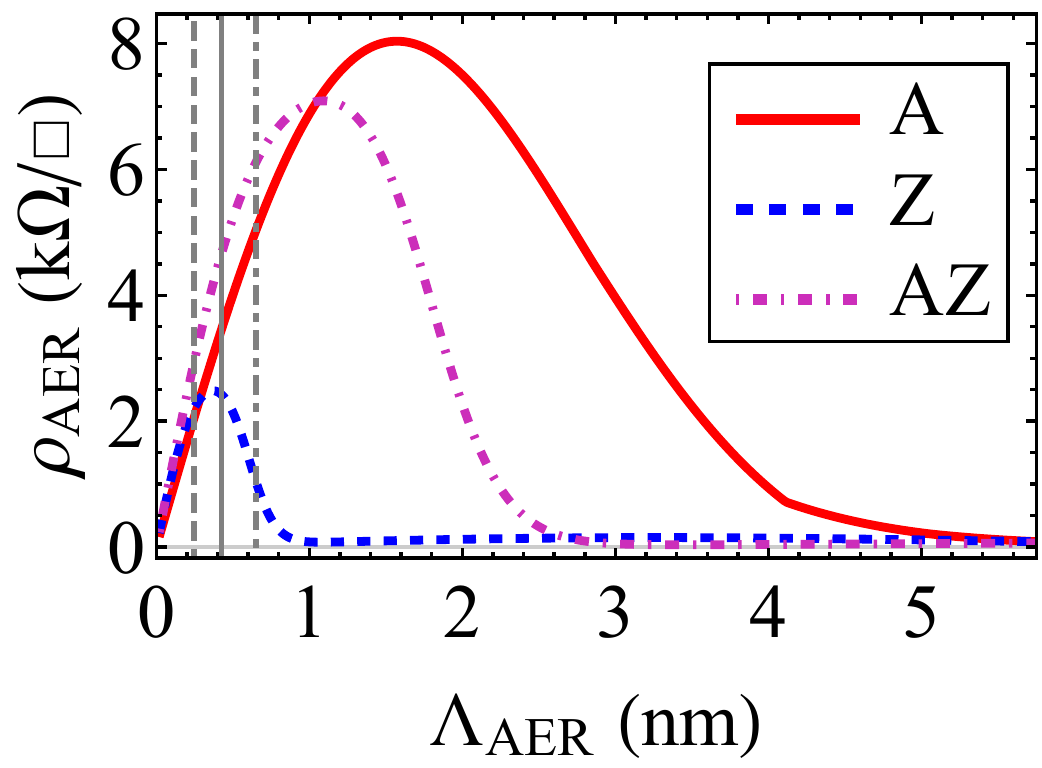}}
	\hspace{0.01\linewidth}
	\subfloat[\ \label{fig:GNR_ER_b}]{\includegraphics[height=0.145\linewidth]{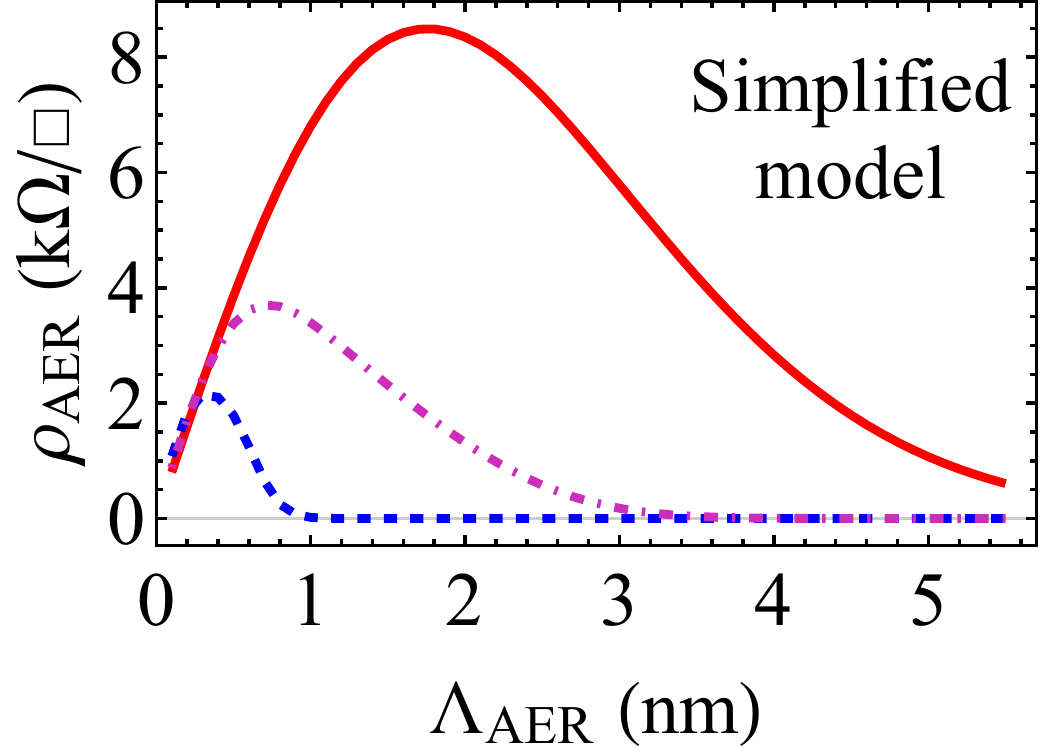}}
	\hspace{0.01\linewidth}
	\subfloat[\ \label{fig:GNR_ER_c}]{\includegraphics[height=0.15\linewidth]{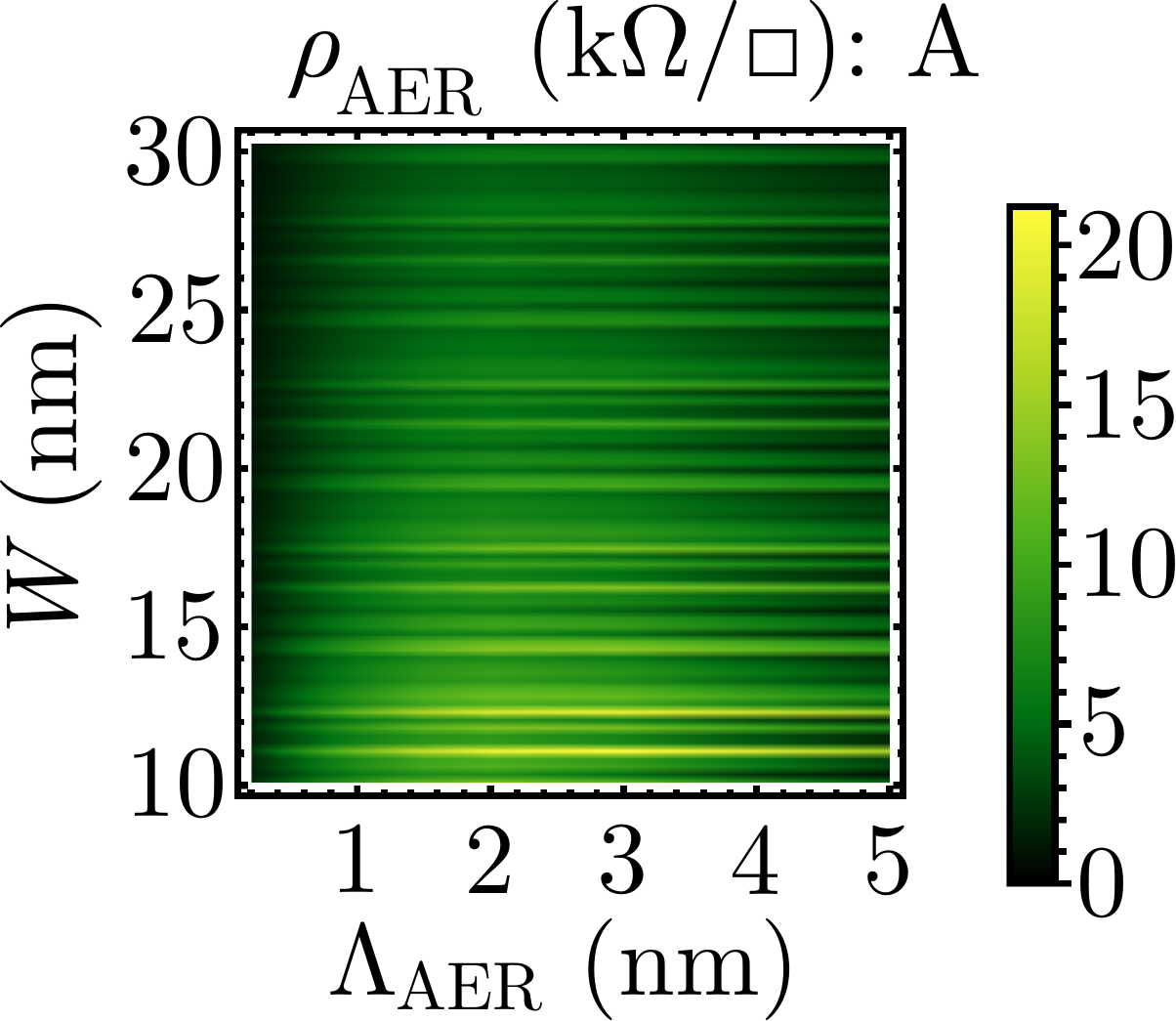}}
	\hspace{0.01\linewidth}
	\subfloat[\ \label{fig:GNR_ER_d}]{\includegraphics[height=0.15\linewidth]{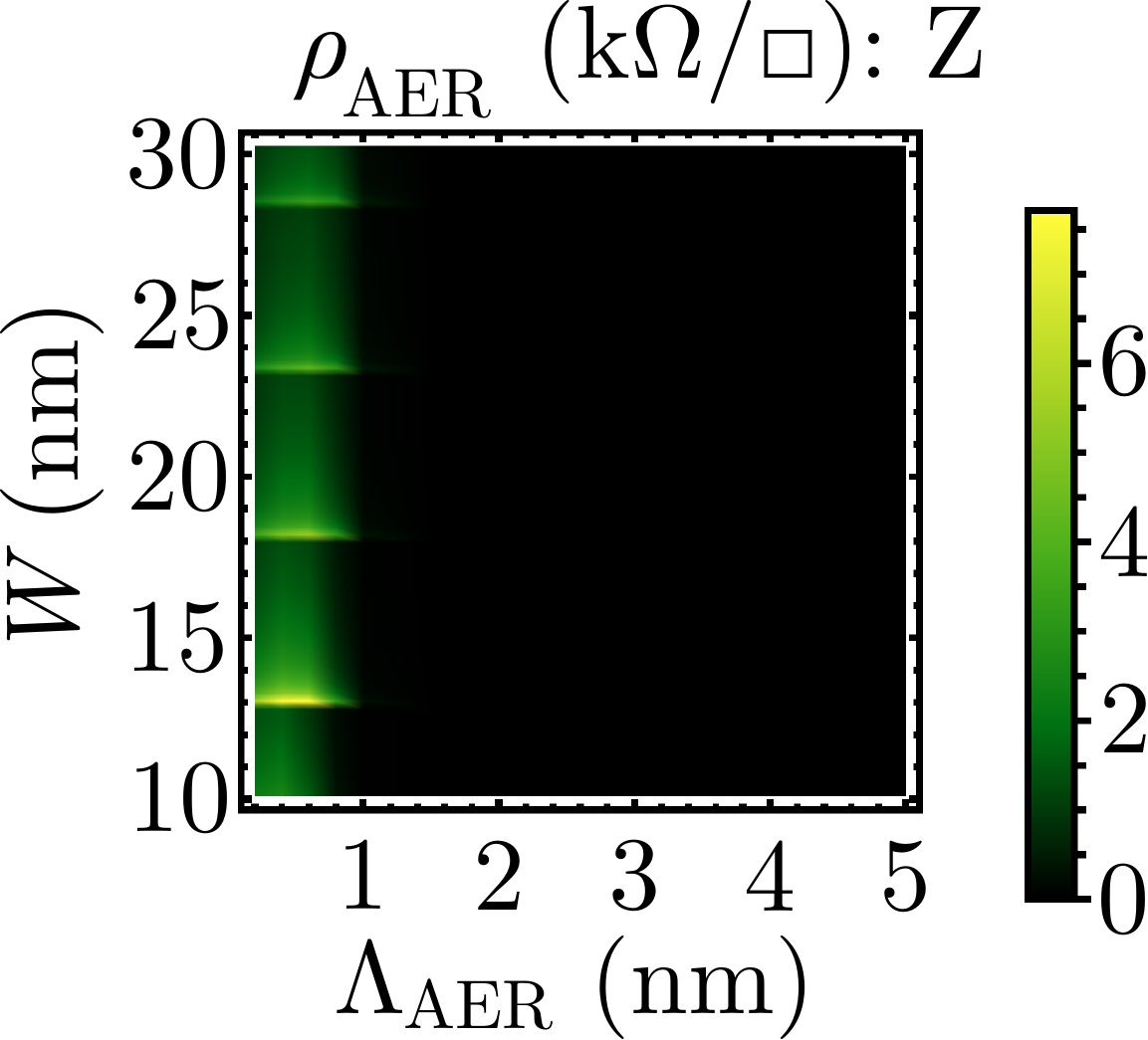}}
	\hspace{0.01\linewidth}
	\subfloat[\ \label{fig:GNR_ER_e}]{\includegraphics[height=0.15\linewidth]{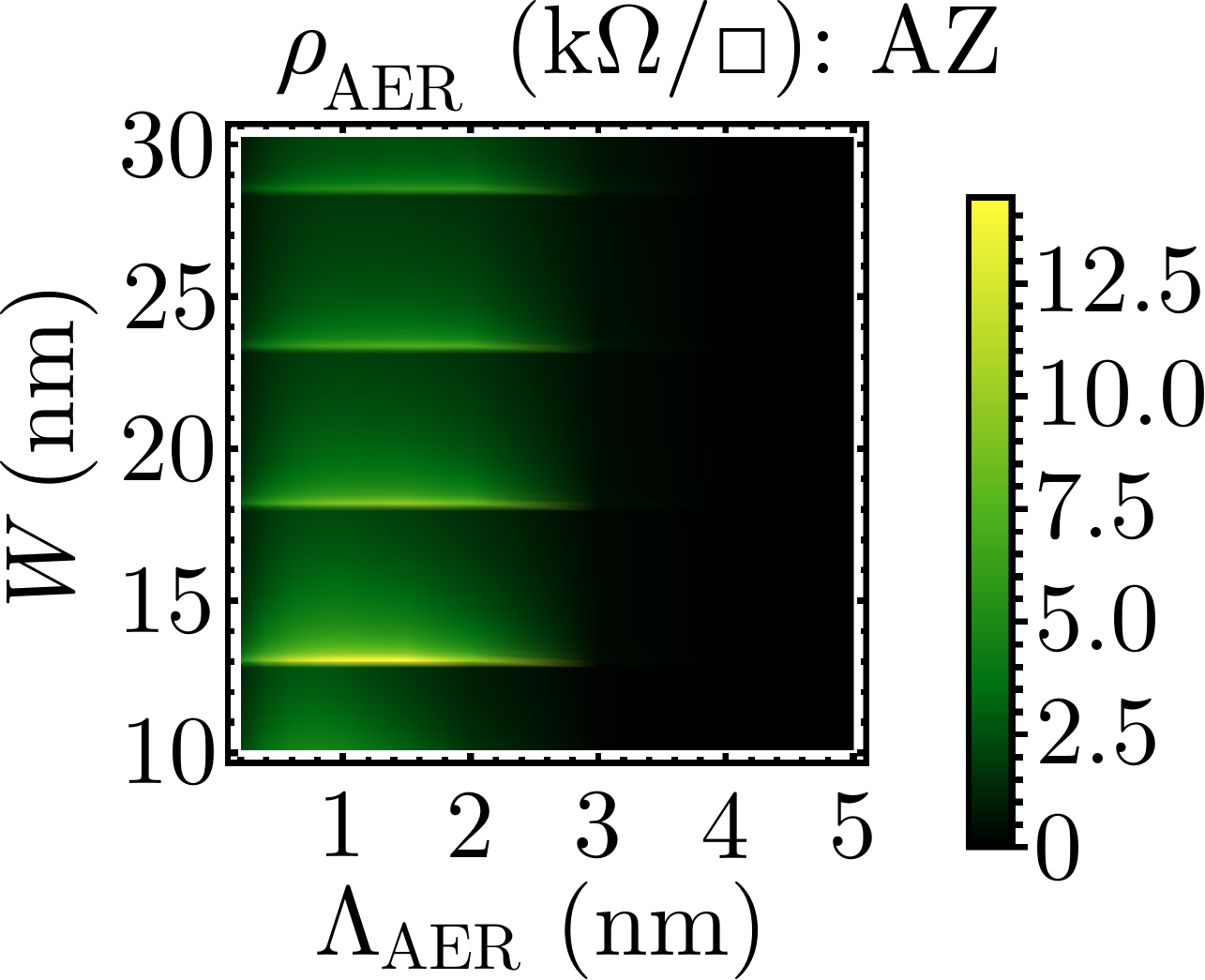}}
	\caption{
		(a),(b) The 2D resistivity $\rho$ due to AER of 10-nm-wide armchair (A), zigzag (Z), and armchair-zigzag (AZ) GNRs as a function of the roughness correlation length with (a) pseudopotential based (see Sec.~\ref{subsec:AER}), (b) simplified scattering rates (see Appendix~\ref{subsec:AER_simplified}). (a) The gray vertical lines in (a) indicate the lengths of the supercell (for each orientation, with matched line type) along the transport direction.
		(c)-(e) The 2D resistivity as a function of the GNR width $W$ and roughness correlation length $\AERCL$ for (c) armchair, (d) zigzag, and (e) armchair-zigzag GNRs.
		(b)-(e) The simplified AER scattering model of Appendix~\ref{subsec:AER_simplified} is considered with $E_\AER = 5$~eV and $\AERSD = \sqrt{3} a_0 / \cos\gamma_\GNR$, corresponding to the width of a single row of carbon atoms.
	}
	\label{fig:GNR_ER}
\end{figure*}

The resistivity as a function of the roughness correlation length $\AERCL$ is shown in Fig.~\subref*{fig:GNR_ER_a} for the three 10-nm-wide GNRs of Fig.~\ref{fig:GNR_Bands}, using the AER scattering rates obtained from Eq.~\eqref{eq:ER_Matrix_El} and considering the Fermi level at $-0.4$~eV (considered throughout this section). Note that the $\AERCL$ has a physical lower limit, related to the carbon atom bond length. For the construction in Eq.~\eqref{eq:potential_AER}, the lower limit is equal to the length of the supercell along the transport direction. Overall, the resistivity is maximal for the armchair GNR and minimal for the zigzag GNR. For all GNR orientations, the resistivity is strongly suppressed for increasing correlation lengths, with the suppression being evident at first in zigzag, then in armchair-zigzag, and finally in armchair GNRs.

From Eq.~\eqref{eq:correlation}, it is clear that the scattering rate between two states is exponentially suppressed when $\AERCL > 2 \sqrt{2} / \Delta k_\parallel$ [see Eq.~\eqref{eq:correlation}]. Considering the wave vector to be equal to the $K$-$K^\prime$ valley separation along the transport direction, $\Delta k_\parallel = \Delta K \sin\gamma_\GNR$ (see simplified GNR model in Appendix~\ref{subsec:GNR_simplified}), we obtain an exponential suppression for the zigzag and armchair-zigzag GNR when $\AERCL > 0.7$~nm and $\AERCL > 1.7$~nm, respectively, which appears to be in good agreement with the exponential suppression of the resistivity.

We observe that the resistivity is exponentially suppressed when intervalley scattering is exponentially suppressed due to the imbalance in the number of forward- and backward-moving channels in each valley. This imbalance originates from the chiral modes which are connected through the flat edge state subbands [see Fig.~\subref*{fig:graphene_d} and Figs.~\subref*{fig:GNR_Bands_b} and \subref{fig:GNR_Bands_c}]. Such a strong suppression has already been theoretically reported as an \textit{anomalous} enhancement of the conductivity of disordered zigzag GNRs, given that intervalley scattering is forbidden or suppressed.\cite{Takane2010} AER scattering satisfies this condition because of its dependence on the wave vector difference. For the armchair GNR, there is no such imbalance and the suppression is therefore minimal. Nonetheless, a suppression of the resistivity eventually appears for increasing correlation length as one should retrieve a perfectly smooth edge and, correspondingly, a vanishing resistivity contribution in the limit $\AERCL \rightarrow +\infty$. In the case that the AER obeys correlation statistics different from Gaussian, the scattering rates are proportional to the Fourier transform of the corresponding two-point correlation function. For exponential correlation statistics for example, we could then expect a power-law suppression of the resistivity when the correlation length exceeds the critical length scale given by $1/\Delta k_\parallel$.

It is important to note that a strong resistivity suppression has not been reported in earlier publications on edge-roughness scattering in graphene ribbons.\cite{Fang2008, Fischetti2011, Goharrizi2011, Xu2012, Dugaev2013}. Our simulation results indicate that this effect only shows up when we calculate the relaxation times self-consistently, considering a subband-quantized ribbon spectrum and a finite edge-roughness correlation length. Furthermore, one frequently simplifies the right-hand side of Eq.~\eqref{eq:BTE} to obtain a closed expression for $\tau_i$, thereby overlooking the self-consistency of the different $\tau$. This appears to be too crude an approximation for this system, because of the strong dependence on the wave vector difference (or subbands) of the highly anisotropic scattering rates. The differences between various approximation schemes for the relaxation times have already been studied in the context of boundary surface roughness scattering metallic nanowires, exhibiting similar behavior.\cite{Moors2014}

Figure~\subref*{fig:GNR_ER_b} shows the resistivity for the same three GNRs as in Fig.~\subref*{fig:GNR_ER_a}, considering the simplified GNR model and scattering rates of Appendix~\ref{appendix:simplified_models}. When considering $E_\AER = 5$~eV (a reasonable value, considering the energy scales of the Kurokawa pseudopotentials) and $\AERSD = \sqrt{3} a_0 / \cos\gamma_\GNR$, being the width of a single row of atoms projected to the direction perpendicular to the GNR transport direction,  the pseudopotential-based results and the simplified model are in excellent agreement. The largest difference is the height of the resistivity peak for the armchair-zigzag GNR, which can easily be improved by adjusting the energy parameter $E_\AER$ appropriately. Figures~\subref*{fig:GNR_ER_c} and \subref{fig:GNR_ER_d} show the simplified model for the same three orientations, considering the same set of parameters, for GNR widths ranging from 10 to 30~nm.

The dependency on the correlation length as in Fig.~\subref*{fig:GNR_ER_b} is retrieved for all widths, while a signature of subband quantization due to confinement is also clearly visible. The highest resistivity peaks are obtained when the top or bottom of the highest occupied subband is just below the Fermi level, offering a large density of states and a strong \textit{catalyst} for intravalley backscattering in the self-consistent relaxation-time solution. In case of an armchair GNR, the peculiar (claromatic) confinement behavior [see Eq.~\eqref{eq:armchair_subbands}] induces a behavior which is highly sensitive to the GNR width, unlike for the other orientations.

\begin{figure*}[htb]
	\centering
	\subfloat[\ \label{fig:comparison_a}]{\includegraphics[width=0.35\linewidth]{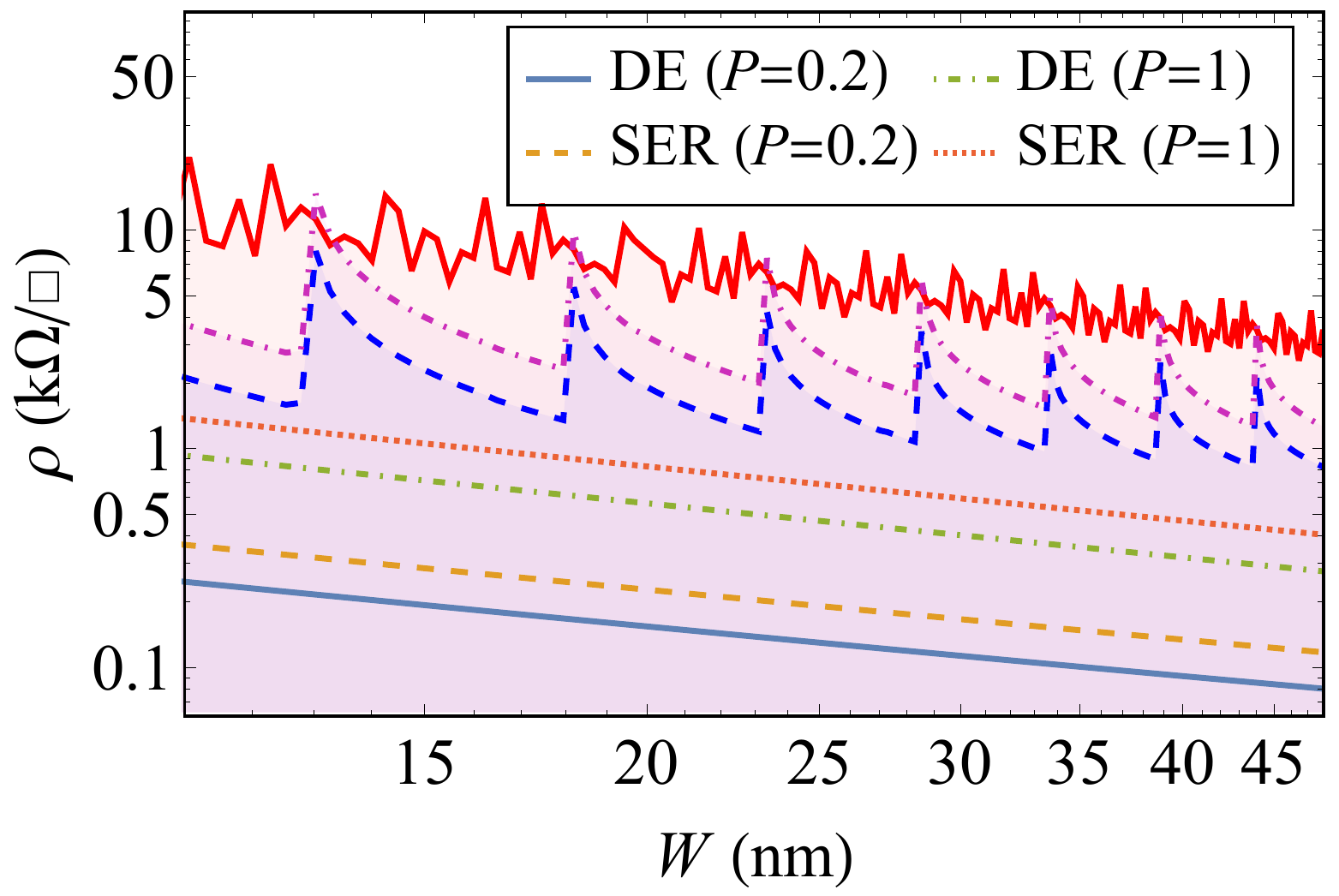}}
	\hspace{0.1\linewidth}
	\subfloat[\ \label{fig:comparison_b}]{\includegraphics[width=0.35\linewidth]{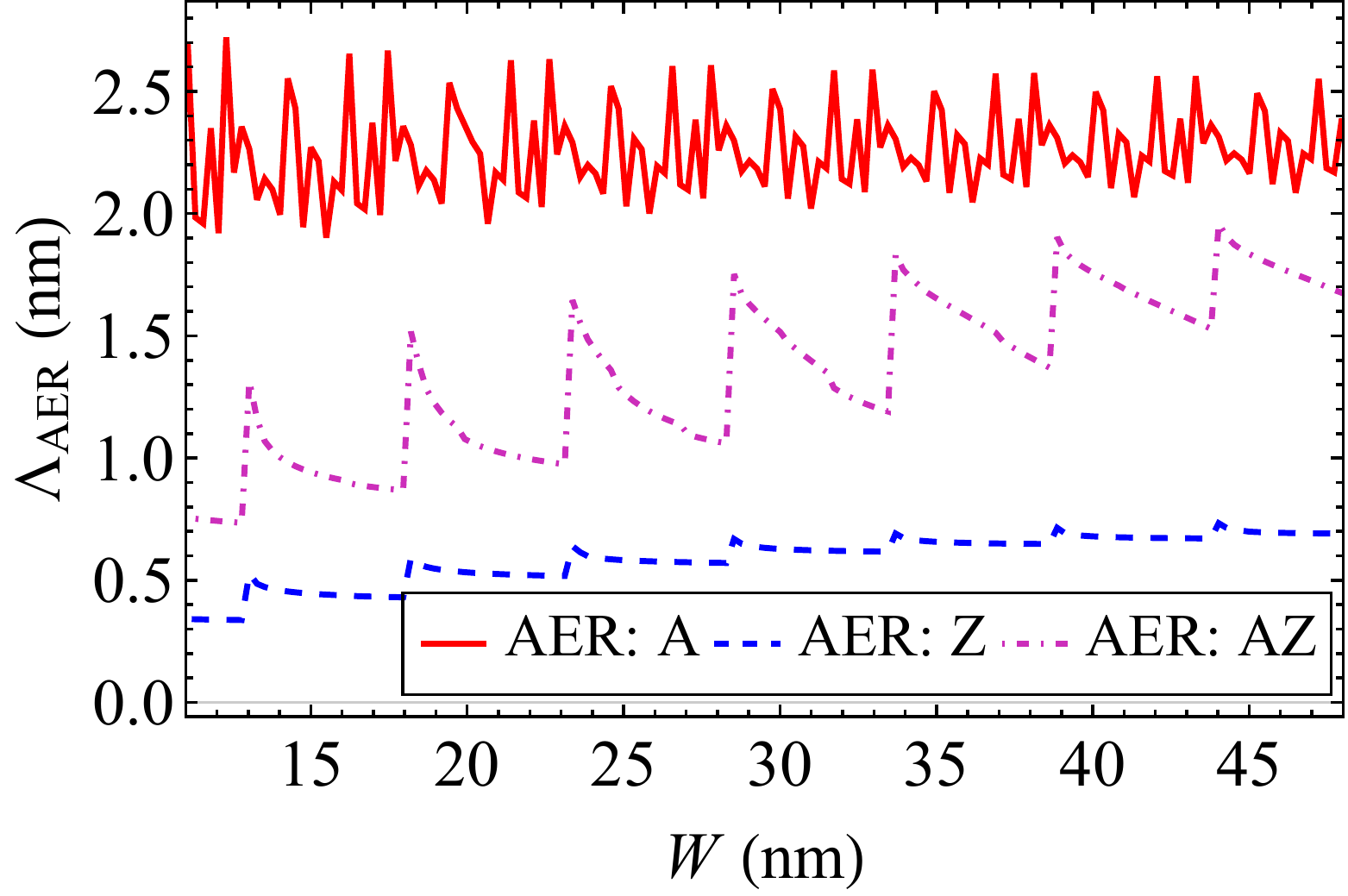}}
	\caption{
		(a) A comparison of the 2D resistivity $\rho$ of GNRs due to DE, SER, and AER scattering [legend in (b) and details on the parameters can be found in the text] as a function of the ribbon width $W$.
		For DE and SER scattering, two values for the probability of DE scattering $P$ are considered.
		For AER scattering, the maximum value of the resistivity, for any value of the correlation length $\AERCL$, is presented. The value of $\AERCL$, for which the resistivity is maximal, is shown in (b) as a function of the GNR width for the three GNR orientations under consideration: armchair (A), zigzag (Z), and armchair-zigzag (AZ).
	}
	\label{fig:comparison}
\end{figure*}

\section{Comparison and discussion}
\label{sec:comparison}
A comparison between the different edge-scattering mechanisms and corresponding models is presented in Fig.~\subref*{fig:comparison_a} for GNRs with average width ranging from 10 to 50~nm. We evaluate the resistivity of the DE and SER scattering models for weakly ($P = 0.2$) and fully diffusive $(P = 1)$ edges and further consider SER with matching standard deviation and correlation length ($\SERCL = \SERSD$) of around 3\% of the total ribbon width ($\SERSD = W/30$). For AER scattering, the maximal resistivity for all correlation lengths is depicted, with the corresponding correlation length for each orientation shown as a function of the width in Fig.~\subref*{fig:comparison_b}. On one hand, this maximum, for edge variations of at most one row of atoms, exceeds the two other types of edge scattering. On the other hand, the contribution is exponentially suppressed when the actual correlation length significantly exceeds the value for which this maximum is reached. In general, the armchair and zigzag GNRs offer a good upper and lower bound for the resistivity of a GNR with arbitrary orientation, respectively. Near the widths for which the top or bottom of the highest occupied subband touches the Fermi level ($-0.4$~eV being considered), the resistivities of the GNRs with different orientation nearly collapse, as backscattering is dominated by intravalley scattering via this particular subband, a process which does not depend on the valley separation in reciprocal space.
On average, all the different edge-scattering mechanisms approximately follow the same scaling behavior, $\rho \propto W^{-\alpha}$, with $\alpha \approx 0.8$.

Note that a specific value needs to be considered for the bulk MFP,  $\lambda_\mathrm{2D}$, in the DE and SER scattering models. This value represents the MFP of electrons moving in a graphene ribbon without considering any edge scattering. As the focus lies on edge scattering in this paper, we have considered a lower limit of the bulk MFP here, only accounting for the dominant electron-phonon interactions (the longitudinal modes of acoustic phonons). The bulk MFP can then be estimated by $\lambda_\mathrm{2D} = 4 \beta \rho_m (\hbar \vF)^3 \upsilon_\mathrm{ph}^2 / (D_\mathrm{ac}^2 \EF)$,\cite{Hwang2008B,DasSarma2011,Fischetti2013} with $\beta \equiv 1/(\kB T)$ (considering room temperature here, $T = 300$~K), $\rho_\mathrm{m} \approx 7 \times 10^{-7}$~kg/m${}^2$ the mass density of graphene, $\upsilon_\mathrm{ph} \approx 2 \times 10^4$~m/s the phonon group velocity, and $D_\mathrm{ac} = 6.5$~eV the considered acoustic deformation potential constant for graphene, yielding $\lambda_\mathrm{2D} \approx \EF \times 1.8$~$ \mu$m/eV.

The resistivity is maximal for the armchair orientation and minimal for the zigzag orientation due to maximally separated $K$ and $K^\prime$ valleys along the transport direction. Each valley features an imbalance in the number of forward- and backward-moving subband channels, which induces a conductivity increase, showing up in the self-consistent relaxation-time solution. The imbalance and resulting resistivity increase is maximally protected when the valleys are maximally separated. If the correlation length exceeds a critical length scale, which is inversely proportional to the $K$-$K^\prime$ valley separation in reciprocal space, intervalley scattering, and the resistivity contribution due to AER, are strongly suppressed. The smallest critical length is realized by the zigzag orientation and is approximately equal to $\sim$0.7~nm (about three times the zigzag supercell length along the transport direction of the ribbon).

It is difficult to provide general trends or crossover regimes between the different edge-scattering mechanisms, as they all depend on one or several parameters ($P$, $\SERSD$, $\AERCL$, \ldots) that can differ for GNR samples of varying quality or obtained with different growth techniques. Furthermore, the resistivity due to AER scattering can only be obtained numerically from the self-consistent set of equations in Eq.~\eqref{eq:BTE}, which makes it hard to extract the general scaling behavior. Nonetheless, there are interesting general observations.
The maximum of the AER-induced resistivity for single-atom-row width deviations exceeds the fully diffusive limit of the diffusive-edge model for all ribbon orientations for example, even up to ribbon widths of around 50~nm. This implies that the phenomenological parameter $P$ cannot be used to represent and replace the average impact of AER scattering, an interpretation that is however often given. Clearly, there are physical values of $\AERCL$ for which a $p(\AERCL)$, which would approximately recover the AER-induced resistivity through \eqref{eq:lambda_diff}, cannot be found, even when neglecting the effect of subband quantization.
Another general observation is that the subband quantization induces resistivity peaks in nonarmchair GNRS as a function of the ribbon width with a periodicity of around 5~nm. We can expect that these peaks are washed out as soon as the width variations become comparable to or exceed this length scale.

\section{Conclusions and outlook}
\label{sec:conclusion}
In this paper, we have presented and compared different models for diffusive transport of electric charge carriers in graphene nanoribbons due to edge scattering. For wide graphene ribbons, the bulk spectrum of 2D graphene can safely be considered and edge scattering is well described with a phenomenological parameter, namely the probability for a diffusive rather than specular scattering event after colliding with a ribbon edge. This approach leads to an analytical expression for the resulting mean free path and can also be extended to account for ribbons with large width variations (much larger than the bond length between neighboring atoms), which we have dubbed structural edge roughness. Structural edge roughness can be characterized by the standard deviation and correlation length of the edge profile. As can be expected, the decrease of the overall mean free path is more pronounced for larger structural edge roughness.

For graphene nanoribbons, the ribbon orientation with respect to the honeycomb lattice, its resulting edge configuration, and the confinement-induced subband quantization also play a role. To model this, we have introduced an atomistic approach based on empirical pseudopotentials. This approach has also been used to obtain a scattering potential for edge roughness on the atomic scale, dubbed atomistic edge roughness. We have combined this approach with a statistical treatment of the edge-roughness properties as well as the Boltzmann transport equation and the self-consistent relaxation-time approximation to obtain an accurate prediction of the resistivity.

The pseudopotential-based model for graphene nanoribbons is well approximated by a simplified description based on the Dirac equation with appropriate boundary condition for the two sublattices. Furthermore, the scattering rates for atomistic edge roughness can be well approximated by considering simplified matrix elements which capture the essential dependence on the wave vector difference. The resulting simplified model can be fitted to the pseudopotential-based model with quantitative agreement and it is easily evaluated for different widths and ribbon orientations. The model nicely demonstrates a strong width dependence of the resistivity for the armchair nanoribbons due to their \textit{claromatic} behavior. Graphene nanoribbons with any other orientation qualitatively behave like a zigzag nanoribbon but show an overall increased resistivity for decreased valley separation. The resistivity of these nanoribbons shows a strong peak when the bottom of a subband crosses the Fermi level.

In the case of heavily-suppressed atomistic edge-roughness scattering due to an optimal combination of the edge-roughness statistics and the valley separation along the transport direction of a graphene nanoribbon, other scattering mechanisms such as local defect, electron-phonon, or remote Coulomb scattering will determine the residual resistivity. These effects are not expected to be strongly width dependent as is the case for edge scattering. Therefore, the edge scattering-suppressed regime, most pronounced for the zigzag orientation, seems to be very promising for nanoscaled device applications that require a high electric current density and a low resistance. 

Our modeling approach for atomistic edge roughness can easily be used to simulate different realistic parameter sets, corresponding to graphene nanoribbons of different very-large-scale integration schemes, and can also be adapted for other two-dimensional materials.

\begin{acknowledgments}
We acknowledge the Research Foundation Flanders (FWO) for supporting K.M.'s research visit at the University of Texas at Dallas, as well as the support by the National Research Fund Luxembourg (FNR) with ATTRACT Grant No.\ 7556175.

\end{acknowledgments}

\bibliography{2018_Moors_Edge_arxiv_v2}{}

\begin{thebibliography}{67}%
\makeatletter
\providecommand \@ifxundefined [1]{%
 \@ifx{#1\undefined}
}%
\providecommand \@ifnum [1]{%
 \ifnum #1\expandafter \@firstoftwo
 \else \expandafter \@secondoftwo
 \fi
}%
\providecommand \@ifx [1]{%
 \ifx #1\expandafter \@firstoftwo
 \else \expandafter \@secondoftwo
 \fi
}%
\providecommand \natexlab [1]{#1}%
\providecommand \enquote  [1]{``#1''}%
\providecommand \bibnamefont  [1]{#1}%
\providecommand \bibfnamefont [1]{#1}%
\providecommand \citenamefont [1]{#1}%
\providecommand \href@noop [0]{\@secondoftwo}%
\providecommand \href [0]{\begingroup \@sanitize@url \@href}%
\providecommand \@href[1]{\@@startlink{#1}\@@href}%
\providecommand \@@href[1]{\endgroup#1\@@endlink}%
\providecommand \@sanitize@url [0]{\catcode `\\12\catcode `\$12\catcode
  `\&12\catcode `\#12\catcode `\^12\catcode `\_12\catcode `\%12\relax}%
\providecommand \@@startlink[1]{}%
\providecommand \@@endlink[0]{}%
\providecommand \url  [0]{\begingroup\@sanitize@url \@url }%
\providecommand \@url [1]{\endgroup\@href {#1}{\urlprefix }}%
\providecommand \urlprefix  [0]{URL }%
\providecommand \Eprint [0]{\href }%
\providecommand \doibase [0]{http://dx.doi.org/}%
\providecommand \selectlanguage [0]{\@gobble}%
\providecommand \bibinfo  [0]{\@secondoftwo}%
\providecommand \bibfield  [0]{\@secondoftwo}%
\providecommand \translation [1]{[#1]}%
\providecommand \BibitemOpen [0]{}%
\providecommand \bibitemStop [0]{}%
\providecommand \bibitemNoStop [0]{.\EOS\space}%
\providecommand \EOS [0]{\spacefactor3000\relax}%
\providecommand \BibitemShut  [1]{\csname bibitem#1\endcsname}%
\let\auto@bib@innerbib\@empty
\bibitem [{\citenamefont {Fujita}\ \emph {et~al.}(1996)\citenamefont {Fujita},
  \citenamefont {Wakabayashi}, \citenamefont {Nakada},\ and\ \citenamefont
  {Kusakabe}}]{Fujita1996}%
  \BibitemOpen
  \bibfield  {author} {\bibinfo {author} {\bibfnamefont {M.}~\bibnamefont
  {Fujita}}, \bibinfo {author} {\bibfnamefont {K.}~\bibnamefont {Wakabayashi}},
  \bibinfo {author} {\bibfnamefont {K.}~\bibnamefont {Nakada}}, \ and\ \bibinfo
  {author} {\bibfnamefont {K.}~\bibnamefont {Kusakabe}},\ }\href {\doibase
  10.1143/jpsj.65.1920} {\bibfield  {journal} {\bibinfo  {journal} {J. Phys.
  Soc. Jpn.}\ }\textbf {\bibinfo {volume} {65}},\ \bibinfo {pages} {1920}
  (\bibinfo {year} {1996})}\BibitemShut {NoStop}%
\bibitem [{\citenamefont {Nakada}\ \emph {et~al.}(1996)\citenamefont {Nakada},
  \citenamefont {Fujita}, \citenamefont {Dresselhaus},\ and\ \citenamefont
  {Dresselhaus}}]{Nakada1996}%
  \BibitemOpen
  \bibfield  {author} {\bibinfo {author} {\bibfnamefont {K.}~\bibnamefont
  {Nakada}}, \bibinfo {author} {\bibfnamefont {M.}~\bibnamefont {Fujita}},
  \bibinfo {author} {\bibfnamefont {G.}~\bibnamefont {Dresselhaus}}, \ and\
  \bibinfo {author} {\bibfnamefont {M.~S.}\ \bibnamefont {Dresselhaus}},\
  }\href {\doibase 10.1103/PhysRevB.54.17954} {\bibfield  {journal} {\bibinfo
  {journal} {Phys. Rev. B}\ }\textbf {\bibinfo {volume} {54}},\ \bibinfo
  {pages} {17954} (\bibinfo {year} {1996})}\BibitemShut {NoStop}%
\bibitem [{\citenamefont {Wakabayashi}\ \emph {et~al.}(1999)\citenamefont
  {Wakabayashi}, \citenamefont {Fujita}, \citenamefont {Ajiki},\ and\
  \citenamefont {Sigrist}}]{Wakabayashi1999}%
  \BibitemOpen
  \bibfield  {author} {\bibinfo {author} {\bibfnamefont {K.}~\bibnamefont
  {Wakabayashi}}, \bibinfo {author} {\bibfnamefont {M.}~\bibnamefont {Fujita}},
  \bibinfo {author} {\bibfnamefont {H.}~\bibnamefont {Ajiki}}, \ and\ \bibinfo
  {author} {\bibfnamefont {M.}~\bibnamefont {Sigrist}},\ }\href {\doibase
  10.1103/PhysRevB.59.8271} {\bibfield  {journal} {\bibinfo  {journal} {Phys.
  Rev. B}\ }\textbf {\bibinfo {volume} {59}},\ \bibinfo {pages} {8271}
  (\bibinfo {year} {1999})}\BibitemShut {NoStop}%
\bibitem [{\citenamefont {Chen}\ \emph {et~al.}(2007)\citenamefont {Chen},
  \citenamefont {Lin}, \citenamefont {Rooks},\ and\ \citenamefont
  {Avouris}}]{Chen2007}%
  \BibitemOpen
  \bibfield  {author} {\bibinfo {author} {\bibfnamefont {Z.}~\bibnamefont
  {Chen}}, \bibinfo {author} {\bibfnamefont {Y.-M.}\ \bibnamefont {Lin}},
  \bibinfo {author} {\bibfnamefont {M.~J.}\ \bibnamefont {Rooks}}, \ and\
  \bibinfo {author} {\bibfnamefont {P.}~\bibnamefont {Avouris}},\ }\href
  {\doibase 10.1016/j.physe.2007.06.020} {\bibfield  {journal} {\bibinfo
  {journal} {Physica E}\ }\textbf {\bibinfo {volume} {40}},\ \bibinfo {pages}
  {228} (\bibinfo {year} {2007})}\BibitemShut {NoStop}%
\bibitem [{\citenamefont {Han}\ \emph {et~al.}(2007)\citenamefont {Han},
  \citenamefont {{\"O}zyilmaz}, \citenamefont {Zhang},\ and\ \citenamefont
  {Kim}}]{Han2007}%
  \BibitemOpen
  \bibfield  {author} {\bibinfo {author} {\bibfnamefont {M.~Y.}\ \bibnamefont
  {Han}}, \bibinfo {author} {\bibfnamefont {B.}~\bibnamefont {{\"O}zyilmaz}},
  \bibinfo {author} {\bibfnamefont {Y.}~\bibnamefont {Zhang}}, \ and\ \bibinfo
  {author} {\bibfnamefont {P.}~\bibnamefont {Kim}},\ }\href {\doibase
  10.1103/PhysRevLett.98.206805} {\bibfield  {journal} {\bibinfo  {journal}
  {Phys. Rev. Lett.}\ }\textbf {\bibinfo {volume} {98}},\ \bibinfo {pages}
  {206805} (\bibinfo {year} {2007})}\BibitemShut {NoStop}%
\bibitem [{\citenamefont {Datta}\ \emph {et~al.}(2008)\citenamefont {Datta},
  \citenamefont {Strachan}, \citenamefont {Khamis},\ and\ \citenamefont
  {Johnson}}]{Datta2008}%
  \BibitemOpen
  \bibfield  {author} {\bibinfo {author} {\bibfnamefont {S.~S.}\ \bibnamefont
  {Datta}}, \bibinfo {author} {\bibfnamefont {D.~R.}\ \bibnamefont {Strachan}},
  \bibinfo {author} {\bibfnamefont {S.~M.}\ \bibnamefont {Khamis}}, \ and\
  \bibinfo {author} {\bibfnamefont {A.~T.~C.}\ \bibnamefont {Johnson}},\ }\href
  {\doibase 10.1021/nl080583r} {\bibfield  {journal} {\bibinfo  {journal} {Nano
  Lett.}\ }\textbf {\bibinfo {volume} {8}},\ \bibinfo {pages} {1912} (\bibinfo
  {year} {2008})}\BibitemShut {NoStop}%
\bibitem [{\citenamefont {Campos-Delgado}\ \emph {et~al.}(2008)\citenamefont
  {Campos-Delgado}, \citenamefont {Romo-Herrera}, \citenamefont {Jia},
  \citenamefont {Cullen}, \citenamefont {Muramatsu}, \citenamefont {Kim},
  \citenamefont {Hayashi}, \citenamefont {Ren}, \citenamefont {Smith},
  \citenamefont {Okuno}, \citenamefont {Ohba}, \citenamefont {Kanoh},
  \citenamefont {Kaneko}, \citenamefont {Endo}, \citenamefont {Terrones},
  \citenamefont {Dresselhaus},\ and\ \citenamefont
  {Terrones}}]{Campos-Delgado2008}%
  \BibitemOpen
  \bibfield  {author} {\bibinfo {author} {\bibfnamefont {J.}~\bibnamefont
  {Campos-Delgado}}, \bibinfo {author} {\bibfnamefont {J.~M.}\ \bibnamefont
  {Romo-Herrera}}, \bibinfo {author} {\bibfnamefont {X.}~\bibnamefont {Jia}},
  \bibinfo {author} {\bibfnamefont {D.~A.}\ \bibnamefont {Cullen}}, \bibinfo
  {author} {\bibfnamefont {H.}~\bibnamefont {Muramatsu}}, \bibinfo {author}
  {\bibfnamefont {Y.~A.}\ \bibnamefont {Kim}}, \bibinfo {author} {\bibfnamefont
  {T.}~\bibnamefont {Hayashi}}, \bibinfo {author} {\bibfnamefont
  {Z.}~\bibnamefont {Ren}}, \bibinfo {author} {\bibfnamefont {D.~J.}\
  \bibnamefont {Smith}}, \bibinfo {author} {\bibfnamefont {Y.}~\bibnamefont
  {Okuno}}, \bibinfo {author} {\bibfnamefont {T.}~\bibnamefont {Ohba}},
  \bibinfo {author} {\bibfnamefont {H.}~\bibnamefont {Kanoh}}, \bibinfo
  {author} {\bibfnamefont {K.}~\bibnamefont {Kaneko}}, \bibinfo {author}
  {\bibfnamefont {M.}~\bibnamefont {Endo}}, \bibinfo {author} {\bibfnamefont
  {H.}~\bibnamefont {Terrones}}, \bibinfo {author} {\bibfnamefont {M.~S.}\
  \bibnamefont {Dresselhaus}}, \ and\ \bibinfo {author} {\bibfnamefont
  {M.}~\bibnamefont {Terrones}},\ }\href {\doibase 10.1021/nl801316d}
  {\bibfield  {journal} {\bibinfo  {journal} {Nano Lett.}\ }\textbf {\bibinfo
  {volume} {8}},\ \bibinfo {pages} {2773} (\bibinfo {year} {2008})}\BibitemShut
  {NoStop}%
\bibitem [{\citenamefont {Yang}\ \emph {et~al.}(2008)\citenamefont {Yang},
  \citenamefont {Dou}, \citenamefont {Rouhanipour}, \citenamefont {Zhi},
  \citenamefont {R{\"{a}}der},\ and\ \citenamefont {M{\"{u}}llen}}]{Yang2008}%
  \BibitemOpen
  \bibfield  {author} {\bibinfo {author} {\bibfnamefont {X.}~\bibnamefont
  {Yang}}, \bibinfo {author} {\bibfnamefont {X.}~\bibnamefont {Dou}}, \bibinfo
  {author} {\bibfnamefont {A.}~\bibnamefont {Rouhanipour}}, \bibinfo {author}
  {\bibfnamefont {L.}~\bibnamefont {Zhi}}, \bibinfo {author} {\bibfnamefont
  {H.~J.}\ \bibnamefont {R{\"{a}}der}}, \ and\ \bibinfo {author} {\bibfnamefont
  {K.}~\bibnamefont {M{\"{u}}llen}},\ }\href {\doibase 10.1021/ja710234t}
  {\bibfield  {journal} {\bibinfo  {journal} {J. Am. Chem. Soc.}\ }\textbf
  {\bibinfo {volume} {130}},\ \bibinfo {pages} {4216} (\bibinfo {year}
  {2008})}\BibitemShut {NoStop}%
\bibitem [{\citenamefont {Li}\ \emph {et~al.}(2008)\citenamefont {Li},
  \citenamefont {Wang}, \citenamefont {Zhang}, \citenamefont {Lee},\ and\
  \citenamefont {Dai}}]{Li2008}%
  \BibitemOpen
  \bibfield  {author} {\bibinfo {author} {\bibfnamefont {X.}~\bibnamefont
  {Li}}, \bibinfo {author} {\bibfnamefont {X.}~\bibnamefont {Wang}}, \bibinfo
  {author} {\bibfnamefont {L.}~\bibnamefont {Zhang}}, \bibinfo {author}
  {\bibfnamefont {S.}~\bibnamefont {Lee}}, \ and\ \bibinfo {author}
  {\bibfnamefont {H.}~\bibnamefont {Dai}},\ }\href {\doibase
  10.1126/science.1150878} {\bibfield  {journal} {\bibinfo  {journal}
  {Science}\ }\textbf {\bibinfo {volume} {319}},\ \bibinfo {pages} {1229}
  (\bibinfo {year} {2008})}\BibitemShut {NoStop}%
\bibitem [{\citenamefont {Jiao}\ \emph {et~al.}(2009)\citenamefont {Jiao},
  \citenamefont {Zhang}, \citenamefont {Wang}, \citenamefont {Diankov},\ and\
  \citenamefont {Dai}}]{Jiao2009}%
  \BibitemOpen
  \bibfield  {author} {\bibinfo {author} {\bibfnamefont {L.}~\bibnamefont
  {Jiao}}, \bibinfo {author} {\bibfnamefont {L.}~\bibnamefont {Zhang}},
  \bibinfo {author} {\bibfnamefont {X.}~\bibnamefont {Wang}}, \bibinfo {author}
  {\bibfnamefont {G.}~\bibnamefont {Diankov}}, \ and\ \bibinfo {author}
  {\bibfnamefont {H.}~\bibnamefont {Dai}},\ }\href {\doibase
  10.1038/nature07919} {\bibfield  {journal} {\bibinfo  {journal} {Nature}\
  }\textbf {\bibinfo {volume} {458}},\ \bibinfo {pages} {877} (\bibinfo {year}
  {2009})}\BibitemShut {NoStop}%
\bibitem [{\citenamefont {Kosynkin}\ \emph {et~al.}(2009)\citenamefont
  {Kosynkin}, \citenamefont {Higginbotham}, \citenamefont {Sinitskii},
  \citenamefont {Lomeda}, \citenamefont {Dimiev}, \citenamefont {Price},\ and\
  \citenamefont {Tour}}]{Kosynkin2009}%
  \BibitemOpen
  \bibfield  {author} {\bibinfo {author} {\bibfnamefont {D.~V.}\ \bibnamefont
  {Kosynkin}}, \bibinfo {author} {\bibfnamefont {A.~L.}\ \bibnamefont
  {Higginbotham}}, \bibinfo {author} {\bibfnamefont {A.}~\bibnamefont
  {Sinitskii}}, \bibinfo {author} {\bibfnamefont {J.~R.}\ \bibnamefont
  {Lomeda}}, \bibinfo {author} {\bibfnamefont {A.}~\bibnamefont {Dimiev}},
  \bibinfo {author} {\bibfnamefont {B.~K.}\ \bibnamefont {Price}}, \ and\
  \bibinfo {author} {\bibfnamefont {J.~M.}\ \bibnamefont {Tour}},\ }\href
  {\doibase 10.1038/nature07872} {\bibfield  {journal} {\bibinfo  {journal}
  {Nature}\ }\textbf {\bibinfo {volume} {458}},\ \bibinfo {pages} {872}
  (\bibinfo {year} {2009})}\BibitemShut {NoStop}%
\bibitem [{\citenamefont {Elías}\ \emph {et~al.}(2010)\citenamefont
  {Elías}, \citenamefont {Botello-Méndez}, \citenamefont
  {Meneses-Rodríguez}, \citenamefont {{Jehová González}}, \citenamefont
  {Ramírez-González}, \citenamefont {Ci}, \citenamefont {Muñoz-Sandoval},
  \citenamefont {Ajayan}, \citenamefont {Terrones},\ and\ \citenamefont
  {Terrones}}]{Elias2010}%
  \BibitemOpen
  \bibfield  {author} {\bibinfo {author} {\bibfnamefont {A.~L.}\ \bibnamefont
  {Elías}}, \bibinfo {author} {\bibfnamefont {A.~R.}\ \bibnamefont
  {Botello-Méndez}}, \bibinfo {author} {\bibfnamefont {D.}~\bibnamefont
  {Meneses-Rodríguez}}, \bibinfo {author} {\bibfnamefont {V.}~\bibnamefont
  {{Jehová González}}}, \bibinfo {author} {\bibfnamefont {D.}~\bibnamefont
  {Ramírez-González}}, \bibinfo {author} {\bibfnamefont {L.}~\bibnamefont
  {Ci}}, \bibinfo {author} {\bibfnamefont {E.}~\bibnamefont
  {Muñoz-Sandoval}}, \bibinfo {author} {\bibfnamefont {P.~M.}\ \bibnamefont
  {Ajayan}}, \bibinfo {author} {\bibfnamefont {H.}~\bibnamefont {Terrones}}, \
  and\ \bibinfo {author} {\bibfnamefont {M.}~\bibnamefont {Terrones}},\ }\href
  {\doibase 10.1021/nl901631z} {\bibfield  {journal} {\bibinfo  {journal} {Nano
  Lett.}\ }\textbf {\bibinfo {volume} {10}},\ \bibinfo {pages} {366} (\bibinfo
  {year} {2010})}\BibitemShut {NoStop}%
\bibitem [{\citenamefont {Jiao}\ \emph {et~al.}(2010)\citenamefont {Jiao},
  \citenamefont {Wang}, \citenamefont {Diankov}, \citenamefont {Wang},\ and\
  \citenamefont {Dai}}]{Jiao2010}%
  \BibitemOpen
  \bibfield  {author} {\bibinfo {author} {\bibfnamefont {L.}~\bibnamefont
  {Jiao}}, \bibinfo {author} {\bibfnamefont {X.}~\bibnamefont {Wang}}, \bibinfo
  {author} {\bibfnamefont {G.}~\bibnamefont {Diankov}}, \bibinfo {author}
  {\bibfnamefont {H.}~\bibnamefont {Wang}}, \ and\ \bibinfo {author}
  {\bibfnamefont {H.}~\bibnamefont {Dai}},\ }\href {\doibase
  10.1038/nnano.2010.54} {\bibfield  {journal} {\bibinfo  {journal} {Nat.
  Nanotechnol.}\ }\textbf {\bibinfo {volume} {5}},\ \bibinfo {pages} {321}
  (\bibinfo {year} {2010})}\BibitemShut {NoStop}%
\bibitem [{\citenamefont {Cai}\ \emph {et~al.}(2010)\citenamefont {Cai},
  \citenamefont {Ruffieux}, \citenamefont {Jaafar}, \citenamefont {Bieri},
  \citenamefont {Braun}, \citenamefont {Blankenburg}, \citenamefont {Muoth},
  \citenamefont {Seitsonen}, \citenamefont {Saleh}, \citenamefont {Feng},
  \citenamefont {M{\"{u}}llen},\ and\ \citenamefont {Fasel}}]{Cai2010}%
  \BibitemOpen
  \bibfield  {author} {\bibinfo {author} {\bibfnamefont {J.}~\bibnamefont
  {Cai}}, \bibinfo {author} {\bibfnamefont {P.}~\bibnamefont {Ruffieux}},
  \bibinfo {author} {\bibfnamefont {R.}~\bibnamefont {Jaafar}}, \bibinfo
  {author} {\bibfnamefont {M.}~\bibnamefont {Bieri}}, \bibinfo {author}
  {\bibfnamefont {T.}~\bibnamefont {Braun}}, \bibinfo {author} {\bibfnamefont
  {S.}~\bibnamefont {Blankenburg}}, \bibinfo {author} {\bibfnamefont
  {M.}~\bibnamefont {Muoth}}, \bibinfo {author} {\bibfnamefont {A.~P.}\
  \bibnamefont {Seitsonen}}, \bibinfo {author} {\bibfnamefont {M.}~\bibnamefont
  {Saleh}}, \bibinfo {author} {\bibfnamefont {X.}~\bibnamefont {Feng}},
  \bibinfo {author} {\bibfnamefont {K.}~\bibnamefont {M{\"{u}}llen}}, \ and\
  \bibinfo {author} {\bibfnamefont {R.}~\bibnamefont {Fasel}},\ }\href
  {\doibase 10.1038/nature09211} {\bibfield  {journal} {\bibinfo  {journal}
  {Nature}\ }\textbf {\bibinfo {volume} {466}},\ \bibinfo {pages} {470}
  (\bibinfo {year} {2010})}\BibitemShut {NoStop}%
\bibitem [{\citenamefont {Tongay}\ \emph {et~al.}(2012)\citenamefont {Tongay},
  \citenamefont {Lemaitre}, \citenamefont {Fridmann}, \citenamefont {Hebard},
  \citenamefont {Gila},\ and\ \citenamefont {Appleton}}]{Tongay2012}%
  \BibitemOpen
  \bibfield  {author} {\bibinfo {author} {\bibfnamefont {S.}~\bibnamefont
  {Tongay}}, \bibinfo {author} {\bibfnamefont {M.}~\bibnamefont {Lemaitre}},
  \bibinfo {author} {\bibfnamefont {J.}~\bibnamefont {Fridmann}}, \bibinfo
  {author} {\bibfnamefont {A.~F.}\ \bibnamefont {Hebard}}, \bibinfo {author}
  {\bibfnamefont {B.~P.}\ \bibnamefont {Gila}}, \ and\ \bibinfo {author}
  {\bibfnamefont {B.~R.}\ \bibnamefont {Appleton}},\ }\href {\doibase
  10.1063/1.3682479} {\bibfield  {journal} {\bibinfo  {journal} {Appl. Phys.
  Lett.}\ }\textbf {\bibinfo {volume} {100}},\ \bibinfo {pages} {073501}
  (\bibinfo {year} {2012})}\BibitemShut {NoStop}%
\bibitem [{\citenamefont {Mohanty}\ \emph {et~al.}(2012)\citenamefont
  {Mohanty}, \citenamefont {Moore}, \citenamefont {Xu}, \citenamefont
  {Sreeprasad}, \citenamefont {Nagaraja}, \citenamefont {Rodriguez},\ and\
  \citenamefont {Berry}}]{Mohanty2012}%
  \BibitemOpen
  \bibfield  {author} {\bibinfo {author} {\bibfnamefont {N.}~\bibnamefont
  {Mohanty}}, \bibinfo {author} {\bibfnamefont {D.}~\bibnamefont {Moore}},
  \bibinfo {author} {\bibfnamefont {Z.}~\bibnamefont {Xu}}, \bibinfo {author}
  {\bibfnamefont {T.~S.}\ \bibnamefont {Sreeprasad}}, \bibinfo {author}
  {\bibfnamefont {A.}~\bibnamefont {Nagaraja}}, \bibinfo {author}
  {\bibfnamefont {A.~A.}\ \bibnamefont {Rodriguez}}, \ and\ \bibinfo {author}
  {\bibfnamefont {V.}~\bibnamefont {Berry}},\ }\href {\doibase
  10.1038/ncomms1834} {\bibfield  {journal} {\bibinfo  {journal} {Nat.
  Commun.}\ }\textbf {\bibinfo {volume} {3}},\ \bibinfo {pages} {844} (\bibinfo
  {year} {2012})}\BibitemShut {NoStop}%
\bibitem [{\citenamefont {Naeemi}\ and\ \citenamefont
  {Meindl}(2007)}]{Naeemi2007}%
  \BibitemOpen
  \bibfield  {author} {\bibinfo {author} {\bibfnamefont {A.}~\bibnamefont
  {Naeemi}}\ and\ \bibinfo {author} {\bibfnamefont {J.~D.}\ \bibnamefont
  {Meindl}},\ }\href {\doibase 10.1109/LED.2007.895452} {\bibfield  {journal}
  {\bibinfo  {journal} {IEEE Electron Device Lett.}\ }\textbf {\bibinfo
  {volume} {28}},\ \bibinfo {pages} {428} (\bibinfo {year} {2007})}\BibitemShut
  {NoStop}%
\bibitem [{\citenamefont {Murali}\ \emph {et~al.}(2009)\citenamefont {Murali},
  \citenamefont {Brenner}, \citenamefont {Yang}, \citenamefont {Beck},\ and\
  \citenamefont {Meindl}}]{Murali2009}%
  \BibitemOpen
  \bibfield  {author} {\bibinfo {author} {\bibfnamefont {R.}~\bibnamefont
  {Murali}}, \bibinfo {author} {\bibfnamefont {K.}~\bibnamefont {Brenner}},
  \bibinfo {author} {\bibfnamefont {Y.}~\bibnamefont {Yang}}, \bibinfo {author}
  {\bibfnamefont {T.}~\bibnamefont {Beck}}, \ and\ \bibinfo {author}
  {\bibfnamefont {J.~D.}\ \bibnamefont {Meindl}},\ }\href {\doibase
  10.1109/LED.2009.2020182} {\bibfield  {journal} {\bibinfo  {journal} {IEEE
  Electron Device Lett.}\ }\textbf {\bibinfo {volume} {30}},\ \bibinfo {pages}
  {611} (\bibinfo {year} {2009})}\BibitemShut {NoStop}%
\bibitem [{\citenamefont {Li}\ \emph {et~al.}(2009)\citenamefont {Li},
  \citenamefont {Xu}, \citenamefont {Srivastava},\ and\ \citenamefont
  {Banerjee}}]{Li2009}%
  \BibitemOpen
  \bibfield  {author} {\bibinfo {author} {\bibfnamefont {H.}~\bibnamefont
  {Li}}, \bibinfo {author} {\bibfnamefont {C.}~\bibnamefont {Xu}}, \bibinfo
  {author} {\bibfnamefont {N.}~\bibnamefont {Srivastava}}, \ and\ \bibinfo
  {author} {\bibfnamefont {K.}~\bibnamefont {Banerjee}},\ }\href {\doibase
  10.1109/TED.2009.2026524} {\bibfield  {journal} {\bibinfo  {journal} {IEEE
  Trans. Electron Devices}\ }\textbf {\bibinfo {volume} {56}},\ \bibinfo
  {pages} {1799} (\bibinfo {year} {2009})}\BibitemShut {NoStop}%
\bibitem [{\citenamefont {Chuan}\ \emph {et~al.}(2009)\citenamefont {Chuan},
  \citenamefont {Hong}, \citenamefont {Banerjee}, \citenamefont {Xu},
  \citenamefont {Li},\ and\ \citenamefont {Banerjee}}]{Chuan2009}%
  \BibitemOpen
  \bibfield  {author} {\bibinfo {author} {\bibfnamefont {X.}~\bibnamefont
  {Chuan}}, \bibinfo {author} {\bibfnamefont {L.}~\bibnamefont {Hong}},
  \bibinfo {author} {\bibfnamefont {K.}~\bibnamefont {Banerjee}}, \bibinfo
  {author} {\bibfnamefont {C.}~\bibnamefont {Xu}}, \bibinfo {author}
  {\bibfnamefont {H.}~\bibnamefont {Li}}, \ and\ \bibinfo {author}
  {\bibfnamefont {K.}~\bibnamefont {Banerjee}},\ }\href {\doibase
  10.1109/TED.2009.2024254} {\bibfield  {journal} {\bibinfo  {journal} {IEEE
  Trans. Electron Devices}\ }\textbf {\bibinfo {volume} {56}},\ \bibinfo
  {pages} {1567} (\bibinfo {year} {2009})}\BibitemShut {NoStop}%
\bibitem [{\citenamefont {Rakheja}\ \emph {et~al.}(2013)\citenamefont
  {Rakheja}, \citenamefont {Kumar},\ and\ \citenamefont
  {Naeemi}}]{Rakheja2013}%
  \BibitemOpen
  \bibfield  {author} {\bibinfo {author} {\bibfnamefont {S.}~\bibnamefont
  {Rakheja}}, \bibinfo {author} {\bibfnamefont {V.}~\bibnamefont {Kumar}}, \
  and\ \bibinfo {author} {\bibfnamefont {A.}~\bibnamefont {Naeemi}},\ }\href
  {\doibase 10.1109/JPROC.2013.2260235} {\bibfield  {journal} {\bibinfo
  {journal} {Proc. IEEE}\ }\textbf {\bibinfo {volume} {101}},\ \bibinfo {pages}
  {1740} (\bibinfo {year} {2013})}\BibitemShut {NoStop}%
\bibitem [{\citenamefont {Cresti}\ \emph {et~al.}(2008)\citenamefont {Cresti},
  \citenamefont {Nemec}, \citenamefont {Biel}, \citenamefont {Niebler},
  \citenamefont {Triozon}, \citenamefont {Cuniberti},\ and\ \citenamefont
  {Roche}}]{Cresti2008}%
  \BibitemOpen
  \bibfield  {author} {\bibinfo {author} {\bibfnamefont {A.}~\bibnamefont
  {Cresti}}, \bibinfo {author} {\bibfnamefont {N.}~\bibnamefont {Nemec}},
  \bibinfo {author} {\bibfnamefont {B.}~\bibnamefont {Biel}}, \bibinfo {author}
  {\bibfnamefont {G.}~\bibnamefont {Niebler}}, \bibinfo {author} {\bibfnamefont
  {F.}~\bibnamefont {Triozon}}, \bibinfo {author} {\bibfnamefont
  {G.}~\bibnamefont {Cuniberti}}, \ and\ \bibinfo {author} {\bibfnamefont
  {S.}~\bibnamefont {Roche}},\ }\href {\doibase 10.1007/s12274-008-8043-2}
  {\bibfield  {journal} {\bibinfo  {journal} {Nano Res.}\ }\textbf {\bibinfo
  {volume} {1}},\ \bibinfo {pages} {361} (\bibinfo {year} {2008})}\BibitemShut
  {NoStop}%
\bibitem [{\citenamefont {Fang}\ \emph {et~al.}(2008)\citenamefont {Fang},
  \citenamefont {Konar}, \citenamefont {Xing},\ and\ \citenamefont
  {Jena}}]{Fang2008}%
  \BibitemOpen
  \bibfield  {author} {\bibinfo {author} {\bibfnamefont {T.}~\bibnamefont
  {Fang}}, \bibinfo {author} {\bibfnamefont {A.}~\bibnamefont {Konar}},
  \bibinfo {author} {\bibfnamefont {H.}~\bibnamefont {Xing}}, \ and\ \bibinfo
  {author} {\bibfnamefont {D.}~\bibnamefont {Jena}},\ }\href {\doibase
  10.1103/PhysRevB.78.205403} {\bibfield  {journal} {\bibinfo  {journal} {Phys.
  Rev. B}\ }\textbf {\bibinfo {volume} {78}},\ \bibinfo {pages} {205403}
  (\bibinfo {year} {2008})}\BibitemShut {NoStop}%
\bibitem [{\citenamefont {Cresti}\ and\ \citenamefont
  {Roche}(2009)}]{Cresti2009}%
  \BibitemOpen
  \bibfield  {author} {\bibinfo {author} {\bibfnamefont {A.}~\bibnamefont
  {Cresti}}\ and\ \bibinfo {author} {\bibfnamefont {S.}~\bibnamefont {Roche}},\
  }\href {\doibase 10.1103/PhysRevB.79.233404} {\bibfield  {journal} {\bibinfo
  {journal} {Phys. Rev. B}\ }\textbf {\bibinfo {volume} {79}},\ \bibinfo
  {pages} {233404} (\bibinfo {year} {2009})}\BibitemShut {NoStop}%
\bibitem [{\citenamefont {Bresciani}\ \emph {et~al.}(2010)\citenamefont
  {Bresciani}, \citenamefont {Palestri}, \citenamefont {Esseni},\ and\
  \citenamefont {Selmi}}]{Bresciani2010}%
  \BibitemOpen
  \bibfield  {author} {\bibinfo {author} {\bibfnamefont {M.}~\bibnamefont
  {Bresciani}}, \bibinfo {author} {\bibfnamefont {P.}~\bibnamefont {Palestri}},
  \bibinfo {author} {\bibfnamefont {D.}~\bibnamefont {Esseni}}, \ and\ \bibinfo
  {author} {\bibfnamefont {L.}~\bibnamefont {Selmi}},\ }\href {\doibase
  10.1016/j.sse.2010.04.038} {\bibfield  {journal} {\bibinfo  {journal}
  {Solid-State Electron.}\ }\textbf {\bibinfo {volume} {54}},\ \bibinfo {pages}
  {1015} (\bibinfo {year} {2010})}\BibitemShut {NoStop}%
\bibitem [{\citenamefont {Takane}(2010)}]{Takane2010}%
  \BibitemOpen
  \bibfield  {author} {\bibinfo {author} {\bibfnamefont {Y.}~\bibnamefont
  {Takane}},\ }\href {\doibase 10.1143/JPSJ.79.024711} {\bibfield  {journal}
  {\bibinfo  {journal} {J. Phys. Soc. Jpn.}\ }\textbf {\bibinfo {volume}
  {79}},\ \bibinfo {pages} {1} (\bibinfo {year} {2010})}\BibitemShut {NoStop}%
\bibitem [{\citenamefont {Fischetti}\ and\ \citenamefont
  {Narayanan}(2011)}]{Fischetti2011}%
  \BibitemOpen
  \bibfield  {author} {\bibinfo {author} {\bibfnamefont {M.~V.}\ \bibnamefont
  {Fischetti}}\ and\ \bibinfo {author} {\bibfnamefont {S.}~\bibnamefont
  {Narayanan}},\ }\href {\doibase 10.1063/1.3650249} {\bibfield  {journal}
  {\bibinfo  {journal} {J. Appl. Phys.}\ }\textbf {\bibinfo {volume} {110}},\
  \bibinfo {pages} {083713} (\bibinfo {year} {2011})}\BibitemShut {NoStop}%
\bibitem [{\citenamefont {Xu}\ and\ \citenamefont {Heinzel}(2012)}]{Xu2012}%
  \BibitemOpen
  \bibfield  {author} {\bibinfo {author} {\bibfnamefont {H.}~\bibnamefont
  {Xu}}\ and\ \bibinfo {author} {\bibfnamefont {T.}~\bibnamefont {Heinzel}},\
  }\href {\doibase 10.1088/0953-8984/24/45/455303} {\bibfield  {journal}
  {\bibinfo  {journal} {J. Phys.: Condens. Matter}\ }\textbf {\bibinfo {volume}
  {24}},\ \bibinfo {pages} {455303} (\bibinfo {year} {2012})}\BibitemShut
  {NoStop}%
\bibitem [{\citenamefont {Fischetti}\ \emph {et~al.}(2013)\citenamefont
  {Fischetti}, \citenamefont {Kim}, \citenamefont {Narayanan}, \citenamefont
  {Ong}, \citenamefont {Sachs}, \citenamefont {Ferry},\ and\ \citenamefont
  {Aboud}}]{Fischetti2013}%
  \BibitemOpen
  \bibfield  {author} {\bibinfo {author} {\bibfnamefont {M.~V.}\ \bibnamefont
  {Fischetti}}, \bibinfo {author} {\bibfnamefont {J.}~\bibnamefont {Kim}},
  \bibinfo {author} {\bibfnamefont {S.}~\bibnamefont {Narayanan}}, \bibinfo
  {author} {\bibfnamefont {Z.-Y.}\ \bibnamefont {Ong}}, \bibinfo {author}
  {\bibfnamefont {C.}~\bibnamefont {Sachs}}, \bibinfo {author} {\bibfnamefont
  {D.~K.}\ \bibnamefont {Ferry}}, \ and\ \bibinfo {author} {\bibfnamefont
  {S.~J.}\ \bibnamefont {Aboud}},\ }\href {\doibase
  10.1088/0953-8984/25/47/473202} {\bibfield  {journal} {\bibinfo  {journal}
  {J. Phys.: Condens. Matter}\ }\textbf {\bibinfo {volume} {25}},\ \bibinfo
  {pages} {473202} (\bibinfo {year} {2013})}\BibitemShut {NoStop}%
\bibitem [{\citenamefont {Dugaev}\ and\ \citenamefont
  {Katsnelson}(2013)}]{Dugaev2013}%
  \BibitemOpen
  \bibfield  {author} {\bibinfo {author} {\bibfnamefont {V.~K.}\ \bibnamefont
  {Dugaev}}\ and\ \bibinfo {author} {\bibfnamefont {M.~I.}\ \bibnamefont
  {Katsnelson}},\ }\href {\doibase 10.1103/PhysRevB.88.235432} {\bibfield
  {journal} {\bibinfo  {journal} {Phys. Rev. B}\ }\textbf {\bibinfo {volume}
  {88}},\ \bibinfo {pages} {235432} (\bibinfo {year} {2013})}\BibitemShut
  {NoStop}%
\bibitem [{\citenamefont {Misawa}\ \emph {et~al.}(2015)\citenamefont {Misawa},
  \citenamefont {Okanaga}, \citenamefont {Mohamad}, \citenamefont {Sakai},\
  and\ \citenamefont {Awano}}]{Misawa2015}%
  \BibitemOpen
  \bibfield  {author} {\bibinfo {author} {\bibfnamefont {T.}~\bibnamefont
  {Misawa}}, \bibinfo {author} {\bibfnamefont {T.}~\bibnamefont {Okanaga}},
  \bibinfo {author} {\bibfnamefont {A.}~\bibnamefont {Mohamad}}, \bibinfo
  {author} {\bibfnamefont {T.}~\bibnamefont {Sakai}}, \ and\ \bibinfo {author}
  {\bibfnamefont {Y.}~\bibnamefont {Awano}},\ }\href {\doibase
  10.7567/JJAP.54.05EB01} {\bibfield  {journal} {\bibinfo  {journal} {Jpn. J.
  Appl. Phys}\ }\textbf {\bibinfo {volume} {54}},\ \bibinfo {pages} {05EB01}
  (\bibinfo {year} {2015})}\BibitemShut {NoStop}%
\bibitem [{\citenamefont {{Contino}}\ \emph {et~al.}(2018)\citenamefont
  {{Contino}}, \citenamefont {{Ciofi}}, \citenamefont {{Wu}}, \citenamefont
  {{Asselberghs}}, \citenamefont {{Celano}}, \citenamefont {{Wilson}},
  \citenamefont {{Tokei}}, \citenamefont {{Groeseneken}},\ and\ \citenamefont
  {{Soree}}}]{Contino2018}%
  \BibitemOpen
  \bibfield  {author} {\bibinfo {author} {\bibfnamefont {A.}~\bibnamefont
  {{Contino}}}, \bibinfo {author} {\bibfnamefont {I.}~\bibnamefont {{Ciofi}}},
  \bibinfo {author} {\bibfnamefont {X.}~\bibnamefont {{Wu}}}, \bibinfo {author}
  {\bibfnamefont {I.}~\bibnamefont {{Asselberghs}}}, \bibinfo {author}
  {\bibfnamefont {U.}~\bibnamefont {{Celano}}}, \bibinfo {author}
  {\bibfnamefont {C.~J.}\ \bibnamefont {{Wilson}}}, \bibinfo {author}
  {\bibfnamefont {Z.}~\bibnamefont {{Tokei}}}, \bibinfo {author} {\bibfnamefont
  {G.}~\bibnamefont {{Groeseneken}}}, \ and\ \bibinfo {author} {\bibfnamefont
  {B.}~\bibnamefont {{Soree}}},\ }\href {\doibase 10.1109/LED.2018.2833633}
  {\bibfield  {journal} {\bibinfo  {journal} {IEEE Electron Device Lett.}\
  }\textbf {\bibinfo {volume} {39}},\ \bibinfo {pages} {1085} (\bibinfo {year}
  {2018})}\BibitemShut {NoStop}%
\bibitem [{\citenamefont {Brey}\ and\ \citenamefont
  {Fertig}(2006{\natexlab{a}})}]{Brey2006B}%
  \BibitemOpen
  \bibfield  {author} {\bibinfo {author} {\bibfnamefont {L.}~\bibnamefont
  {Brey}}\ and\ \bibinfo {author} {\bibfnamefont {H.~A.}\ \bibnamefont
  {Fertig}},\ }\href {\doibase 10.1103/PhysRevB.73.235411} {\bibfield
  {journal} {\bibinfo  {journal} {Phys. Rev. B}\ }\textbf {\bibinfo {volume}
  {73}},\ \bibinfo {pages} {235411} (\bibinfo {year}
  {2006}{\natexlab{a}})}\BibitemShut {NoStop}%
\bibitem [{\citenamefont {Kurokawa}\ \emph {et~al.}(2000)\citenamefont
  {Kurokawa}, \citenamefont {Nomura}, \citenamefont {Takemori},\ and\
  \citenamefont {Aoyagi}}]{Kurokawa2000}%
  \BibitemOpen
  \bibfield  {author} {\bibinfo {author} {\bibfnamefont {Y.}~\bibnamefont
  {Kurokawa}}, \bibinfo {author} {\bibfnamefont {S.}~\bibnamefont {Nomura}},
  \bibinfo {author} {\bibfnamefont {T.}~\bibnamefont {Takemori}}, \ and\
  \bibinfo {author} {\bibfnamefont {Y.}~\bibnamefont {Aoyagi}},\ }\href
  {\doibase 10.1103/PhysRevB.61.12616} {\bibfield  {journal} {\bibinfo
  {journal} {Phys. Rev. B}\ }\textbf {\bibinfo {volume} {61}},\ \bibinfo
  {pages} {12616} (\bibinfo {year} {2000})}\BibitemShut {NoStop}%
\bibitem [{\citenamefont {Gmitra}\ \emph {et~al.}(2009)\citenamefont {Gmitra},
  \citenamefont {Konschuh}, \citenamefont {Ertler}, \citenamefont
  {Ambrosch-Draxl},\ and\ \citenamefont {Fabian}}]{Gmitra2009}%
  \BibitemOpen
  \bibfield  {author} {\bibinfo {author} {\bibfnamefont {M.}~\bibnamefont
  {Gmitra}}, \bibinfo {author} {\bibfnamefont {S.}~\bibnamefont {Konschuh}},
  \bibinfo {author} {\bibfnamefont {C.}~\bibnamefont {Ertler}}, \bibinfo
  {author} {\bibfnamefont {C.}~\bibnamefont {Ambrosch-Draxl}}, \ and\ \bibinfo
  {author} {\bibfnamefont {J.}~\bibnamefont {Fabian}},\ }\href {\doibase
  10.1103/PhysRevB.80.235431} {\bibfield  {journal} {\bibinfo  {journal} {Phys.
  Rev. B}\ }\textbf {\bibinfo {volume} {80}},\ \bibinfo {pages} {235431}
  (\bibinfo {year} {2009})}\BibitemShut {NoStop}%
\bibitem [{\citenamefont {Martin}(2004)}]{Martin2004}%
  \BibitemOpen
  \bibfield  {author} {\bibinfo {author} {\bibfnamefont {R.~M.}\ \bibnamefont
  {Martin}},\ }\href {\doibase 10.1017/CBO9780511805769} {\emph {\bibinfo
  {title} {{Electronic structure: basic theory and practical methods}}}}\
  (\bibinfo  {publisher} {Cambridge University Press},\ \bibinfo {year}
  {2004})\BibitemShut {NoStop}%
\bibitem [{\citenamefont {Frigo}\ and\ \citenamefont
  {Johnson}(2005)}]{Frigo2005}%
  \BibitemOpen
  \bibfield  {author} {\bibinfo {author} {\bibfnamefont {M.}~\bibnamefont
  {Frigo}}\ and\ \bibinfo {author} {\bibfnamefont {S.~G.}\ \bibnamefont
  {Johnson}},\ }\href {\doibase 10.1109/JPROC.2004.840301} {\bibfield
  {journal} {\bibinfo  {journal} {Proc. IEEE}\ }\textbf {\bibinfo {volume}
  {93}},\ \bibinfo {pages} {216} (\bibinfo {year} {2005})}\BibitemShut
  {NoStop}%
\bibitem [{\citenamefont {Kresse}\ and\ \citenamefont
  {Furthm{\"u}ller}(1996)}]{Kresse1996}%
  \BibitemOpen
  \bibfield  {author} {\bibinfo {author} {\bibfnamefont {G.}~\bibnamefont
  {Kresse}}\ and\ \bibinfo {author} {\bibfnamefont {J.}~\bibnamefont
  {Furthm{\"u}ller}},\ }\href {\doibase 10.1103/PhysRevB.54.11169} {\bibfield
  {journal} {\bibinfo  {journal} {Phys. Rev. B}\ }\textbf {\bibinfo {volume}
  {54}},\ \bibinfo {pages} {11169} (\bibinfo {year} {1996})}\BibitemShut
  {NoStop}%
\bibitem [{\citenamefont {{Van de Put}}\ \emph {et~al.}(2016)\citenamefont
  {{Van de Put}}, \citenamefont {Vandenberghe}, \citenamefont {Sor{\'{e}}e},
  \citenamefont {Magnus},\ and\ \citenamefont {Fischetti}}]{VandePut2016}%
  \BibitemOpen
  \bibfield  {author} {\bibinfo {author} {\bibfnamefont {M.~L.}\ \bibnamefont
  {{Van de Put}}}, \bibinfo {author} {\bibfnamefont {W.~G.}\ \bibnamefont
  {Vandenberghe}}, \bibinfo {author} {\bibfnamefont {B.}~\bibnamefont
  {Sor{\'{e}}e}}, \bibinfo {author} {\bibfnamefont {W.}~\bibnamefont {Magnus}},
  \ and\ \bibinfo {author} {\bibfnamefont {M.~V.}\ \bibnamefont {Fischetti}},\
  }\href {\doibase 10.1063/1.4953148} {\bibfield  {journal} {\bibinfo
  {journal} {J. Appl. Phys.}\ }\textbf {\bibinfo {volume} {119}},\ \bibinfo
  {pages} {214306} (\bibinfo {year} {2016})}\BibitemShut {NoStop}%
\bibitem [{\citenamefont {Wood}\ and\ \citenamefont {Zunger}(1985)}]{Wood1985}%
  \BibitemOpen
  \bibfield  {author} {\bibinfo {author} {\bibfnamefont {D.~M.}\ \bibnamefont
  {Wood}}\ and\ \bibinfo {author} {\bibfnamefont {A.}~\bibnamefont {Zunger}},\
  }\href {\doibase 10.1088/0305-4470/18/9/018} {\bibfield  {journal} {\bibinfo
  {journal} {J. Phys. A: Math. Gen.}\ }\textbf {\bibinfo {volume} {18}},\
  \bibinfo {pages} {1343} (\bibinfo {year} {1985})}\BibitemShut {NoStop}%
\bibitem [{\citenamefont {Clar}(1972)}]{Clar1972}%
  \BibitemOpen
  \bibfield  {author} {\bibinfo {author} {\bibfnamefont {E.}~\bibnamefont
  {Clar}},\ }\href@noop {} {\emph {\bibinfo {title} {{The Aromatic Sextet}}}}\
  (\bibinfo  {publisher} {Wiley},\ \bibinfo {year} {1972})\BibitemShut
  {NoStop}%
\bibitem [{\citenamefont {Balaban}\ and\ \citenamefont
  {Klein}(2009)}]{Balaban2009}%
  \BibitemOpen
  \bibfield  {author} {\bibinfo {author} {\bibfnamefont {A.~T.}\ \bibnamefont
  {Balaban}}\ and\ \bibinfo {author} {\bibfnamefont {D.~J.}\ \bibnamefont
  {Klein}},\ }\href {\doibase 10.1021/jp9082618} {\bibfield  {journal}
  {\bibinfo  {journal} {J. Phys. Chem. C}\ }\textbf {\bibinfo {volume} {113}},\
  \bibinfo {pages} {19123} (\bibinfo {year} {2009})}\BibitemShut {NoStop}%
\bibitem [{\citenamefont {Ryu}\ and\ \citenamefont {Hatsugai}(2002)}]{Ryu2002}%
  \BibitemOpen
  \bibfield  {author} {\bibinfo {author} {\bibfnamefont {S.}~\bibnamefont
  {Ryu}}\ and\ \bibinfo {author} {\bibfnamefont {Y.}~\bibnamefont {Hatsugai}},\
  }\href {\doibase 10.1103/PhysRevLett.89.077002} {\bibfield  {journal}
  {\bibinfo  {journal} {Phys. Rev. Lett.}\ }\textbf {\bibinfo {volume} {89}},\
  \bibinfo {pages} {077002} (\bibinfo {year} {2002})}\BibitemShut {NoStop}%
\bibitem [{\citenamefont {Brey}\ and\ \citenamefont
  {Fertig}(2006{\natexlab{b}})}]{Brey2006A}%
  \BibitemOpen
  \bibfield  {author} {\bibinfo {author} {\bibfnamefont {L.}~\bibnamefont
  {Brey}}\ and\ \bibinfo {author} {\bibfnamefont {H.~A.}\ \bibnamefont
  {Fertig}},\ }\href {\doibase 10.1103/PhysRevB.73.195408} {\bibfield
  {journal} {\bibinfo  {journal} {Phys. Rev. B}\ }\textbf {\bibinfo {volume}
  {73}},\ \bibinfo {pages} {195408} (\bibinfo {year}
  {2006}{\natexlab{b}})}\BibitemShut {NoStop}%
\bibitem [{\citenamefont {Akhmerov}\ and\ \citenamefont
  {Beenakker}(2008)}]{Akhmerov2008}%
  \BibitemOpen
  \bibfield  {author} {\bibinfo {author} {\bibfnamefont {A.~R.}\ \bibnamefont
  {Akhmerov}}\ and\ \bibinfo {author} {\bibfnamefont {C.~W.~J.}\ \bibnamefont
  {Beenakker}},\ }\href {\doibase 10.1103/PhysRevB.77.085423} {\bibfield
  {journal} {\bibinfo  {journal} {Phys. Rev. B}\ }\textbf {\bibinfo {volume}
  {77}},\ \bibinfo {pages} {085423} (\bibinfo {year} {2008})}\BibitemShut
  {NoStop}%
\bibitem [{\citenamefont {Delplace}\ \emph {et~al.}(2011)\citenamefont
  {Delplace}, \citenamefont {Ullmo},\ and\ \citenamefont
  {Montambaux}}]{Delplace2011}%
  \BibitemOpen
  \bibfield  {author} {\bibinfo {author} {\bibfnamefont {P.}~\bibnamefont
  {Delplace}}, \bibinfo {author} {\bibfnamefont {D.}~\bibnamefont {Ullmo}}, \
  and\ \bibinfo {author} {\bibfnamefont {G.}~\bibnamefont {Montambaux}},\
  }\href {\doibase 10.1103/PhysRevB.84.195452} {\bibfield  {journal} {\bibinfo
  {journal} {Phys. Rev. B}\ }\textbf {\bibinfo {volume} {84}},\ \bibinfo
  {pages} {195452} (\bibinfo {year} {2011})}\BibitemShut {NoStop}%
\bibitem [{\citenamefont {van Miert}\ \emph {et~al.}(2016)\citenamefont {van
  Miert}, \citenamefont {Ortix},\ and\ \citenamefont {Smith}}]{VanMiert2016}%
  \BibitemOpen
  \bibfield  {author} {\bibinfo {author} {\bibfnamefont {G.}~\bibnamefont {van
  Miert}}, \bibinfo {author} {\bibfnamefont {C.}~\bibnamefont {Ortix}}, \ and\
  \bibinfo {author} {\bibfnamefont {C.~M.}\ \bibnamefont {Smith}},\ }\href
  {\doibase 10.1088/2053-1583/4/1/015023} {\bibfield  {journal} {\bibinfo
  {journal} {2D Mater.}\ }\textbf {\bibinfo {volume} {4}},\ \bibinfo {pages}
  {015023} (\bibinfo {year} {2016})}\BibitemShut {NoStop}%
\bibitem [{\citenamefont {Bhowmick}\ and\ \citenamefont
  {Shenoy}(2008)}]{Bhowmick2008}%
  \BibitemOpen
  \bibfield  {author} {\bibinfo {author} {\bibfnamefont {S.}~\bibnamefont
  {Bhowmick}}\ and\ \bibinfo {author} {\bibfnamefont {V.~B.}\ \bibnamefont
  {Shenoy}},\ }\href {\doibase 10.1063/1.2943678} {\bibfield  {journal}
  {\bibinfo  {journal} {J. Chem. Phys.}\ }\textbf {\bibinfo {volume} {128}},\
  \bibinfo {pages} {244717} (\bibinfo {year} {2008})}\BibitemShut {NoStop}%
\bibitem [{\citenamefont {Jung}\ \emph {et~al.}(2009)\citenamefont {Jung},
  \citenamefont {Pereg-Barnea},\ and\ \citenamefont {MacDonald}}]{Jung2009}%
  \BibitemOpen
  \bibfield  {author} {\bibinfo {author} {\bibfnamefont {J.}~\bibnamefont
  {Jung}}, \bibinfo {author} {\bibfnamefont {T.}~\bibnamefont {Pereg-Barnea}},
  \ and\ \bibinfo {author} {\bibfnamefont {A.~H.}\ \bibnamefont {MacDonald}},\
  }\href {\doibase 10.1103/PhysRevLett.102.227205} {\bibfield  {journal}
  {\bibinfo  {journal} {Phys. Rev. Lett.}\ }\textbf {\bibinfo {volume} {102}},\
  \bibinfo {pages} {227205} (\bibinfo {year} {2009})}\BibitemShut {NoStop}%
\bibitem [{\citenamefont {Magda}\ \emph {et~al.}(2014)\citenamefont {Magda},
  \citenamefont {Jin}, \citenamefont {Hagym{\'a}si}, \citenamefont
  {Vancs{\'o}}, \citenamefont {Osv{\'a}th}, \citenamefont {Nemes-Incze},
  \citenamefont {Hwang}, \citenamefont {Bir{\'o}},\ and\ \citenamefont
  {Tapaszt{\'o}}}]{Magda2014}%
  \BibitemOpen
  \bibfield  {author} {\bibinfo {author} {\bibfnamefont {G.~Z.}\ \bibnamefont
  {Magda}}, \bibinfo {author} {\bibfnamefont {X.}~\bibnamefont {Jin}}, \bibinfo
  {author} {\bibfnamefont {I.}~\bibnamefont {Hagym{\'a}si}}, \bibinfo {author}
  {\bibfnamefont {P.}~\bibnamefont {Vancs{\'o}}}, \bibinfo {author}
  {\bibfnamefont {Z.}~\bibnamefont {Osv{\'a}th}}, \bibinfo {author}
  {\bibfnamefont {P.}~\bibnamefont {Nemes-Incze}}, \bibinfo {author}
  {\bibfnamefont {C.}~\bibnamefont {Hwang}}, \bibinfo {author} {\bibfnamefont
  {L.~P.}\ \bibnamefont {Bir{\'o}}}, \ and\ \bibinfo {author} {\bibfnamefont
  {L.}~\bibnamefont {Tapaszt{\'o}}},\ }\href {\doibase 10.1038/nature13831}
  {\bibfield  {journal} {\bibinfo  {journal} {Nature}\ }\textbf {\bibinfo
  {volume} {514}},\ \bibinfo {pages} {608} (\bibinfo {year}
  {2014})}\BibitemShut {NoStop}%
\bibitem [{\citenamefont {Fuchs}(1938)}]{Fuchs1938}%
  \BibitemOpen
  \bibfield  {author} {\bibinfo {author} {\bibfnamefont {K.}~\bibnamefont
  {Fuchs}},\ }\href {\doibase 10.1017/S0305004100019952} {\bibfield  {journal}
  {\bibinfo  {journal} {Math. Proc. Cambridge Philos. Soc.}\ }\textbf {\bibinfo
  {volume} {34}},\ \bibinfo {pages} {100} (\bibinfo {year} {1938})}\BibitemShut
  {NoStop}%
\bibitem [{\citenamefont {Sondheimer}(1952)}]{Sondheimer1952}%
  \BibitemOpen
  \bibfield  {author} {\bibinfo {author} {\bibfnamefont {E.}~\bibnamefont
  {Sondheimer}},\ }\href {\doibase 10.1080/00018735200101151} {\bibfield
  {journal} {\bibinfo  {journal} {Adv. Phys.}\ }\textbf {\bibinfo {volume}
  {1}},\ \bibinfo {pages} {1} (\bibinfo {year} {1952})}\BibitemShut {NoStop}%
\bibitem [{\citenamefont {Ando}\ \emph {et~al.}(1982)\citenamefont {Ando},
  \citenamefont {Fowler},\ and\ \citenamefont {Stern}}]{Ando1982}%
  \BibitemOpen
  \bibfield  {author} {\bibinfo {author} {\bibfnamefont {T.}~\bibnamefont
  {Ando}}, \bibinfo {author} {\bibfnamefont {A.~B.}\ \bibnamefont {Fowler}}, \
  and\ \bibinfo {author} {\bibfnamefont {F.}~\bibnamefont {Stern}},\ }\href
  {\doibase 10.1103/RevModPhys.54.437} {\bibfield  {journal} {\bibinfo
  {journal} {Rev. Mod. Phys.}\ }\textbf {\bibinfo {volume} {54}},\ \bibinfo
  {pages} {437} (\bibinfo {year} {1982})}\BibitemShut {NoStop}%
\bibitem [{\citenamefont {Goodnick}\ \emph {et~al.}(1985)\citenamefont
  {Goodnick}, \citenamefont {Ferry}, \citenamefont {Wilmsen}, \citenamefont
  {Liliental}, \citenamefont {Fathy},\ and\ \citenamefont
  {Krivanek}}]{Goodnick1985}%
  \BibitemOpen
  \bibfield  {author} {\bibinfo {author} {\bibfnamefont {S.~M.}\ \bibnamefont
  {Goodnick}}, \bibinfo {author} {\bibfnamefont {D.~K.}\ \bibnamefont {Ferry}},
  \bibinfo {author} {\bibfnamefont {C.~W.}\ \bibnamefont {Wilmsen}}, \bibinfo
  {author} {\bibfnamefont {Z.}~\bibnamefont {Liliental}}, \bibinfo {author}
  {\bibfnamefont {D.}~\bibnamefont {Fathy}}, \ and\ \bibinfo {author}
  {\bibfnamefont {O.~L.}\ \bibnamefont {Krivanek}},\ }\href
  {https://link.aps.org/doi/10.1103/PhysRevB.32.8171} {\bibfield  {journal}
  {\bibinfo  {journal} {Phys. Rev. B}\ }\textbf {\bibinfo {volume} {32}},\
  \bibinfo {pages} {8171} (\bibinfo {year} {1985})}\BibitemShut {NoStop}%
\bibitem [{\citenamefont {Yu}\ \emph {et~al.}(2002)\citenamefont {Yu},
  \citenamefont {Vasileska}, \citenamefont {Goodnick}, \citenamefont {Grazul},
  \citenamefont {Green}, \citenamefont {Kim}, \citenamefont {Evans-Lutterodt},
  \citenamefont {Liu}, \citenamefont {Lyding}, \citenamefont {Mansfield},
  \citenamefont {Muller}, \citenamefont {Sorsch}, \citenamefont {Timp},\ and\
  \citenamefont {Timp}}]{Yu2002}%
  \BibitemOpen
  \bibfield  {author} {\bibinfo {author} {\bibfnamefont {J.}~\bibnamefont
  {Yu}}, \bibinfo {author} {\bibfnamefont {D.}~\bibnamefont {Vasileska}},
  \bibinfo {author} {\bibfnamefont {S.}~\bibnamefont {Goodnick}}, \bibinfo
  {author} {\bibfnamefont {J.}~\bibnamefont {Grazul}}, \bibinfo {author}
  {\bibfnamefont {M.}~\bibnamefont {Green}}, \bibinfo {author} {\bibfnamefont
  {C.~Y.}\ \bibnamefont {Kim}}, \bibinfo {author} {\bibfnamefont
  {K.}~\bibnamefont {Evans-Lutterodt}}, \bibinfo {author} {\bibfnamefont
  {L.}~\bibnamefont {Liu}}, \bibinfo {author} {\bibfnamefont {J.}~\bibnamefont
  {Lyding}}, \bibinfo {author} {\bibfnamefont {W.}~\bibnamefont {Mansfield}},
  \bibinfo {author} {\bibfnamefont {D.}~\bibnamefont {Muller}}, \bibinfo
  {author} {\bibfnamefont {T.}~\bibnamefont {Sorsch}}, \bibinfo {author}
  {\bibfnamefont {R.}~\bibnamefont {Timp}}, \ and\ \bibinfo {author}
  {\bibfnamefont {G.}~\bibnamefont {Timp}},\ }in\ \href@noop {} {\emph
  {\bibinfo {booktitle} {Proceedings of the 2002 IEEE Silicon Nanoelectronics
  Workshop}}}\ (\bibinfo  {publisher} {IEEE},\ \bibinfo {address} {New York},\
  \bibinfo {year} {2002})\ p.~\bibinfo {pages} {75}\BibitemShut {NoStop}%
\bibitem [{\citenamefont {Prange}\ and\ \citenamefont
  {Nee}(1968)}]{Prange1968}%
  \BibitemOpen
  \bibfield  {author} {\bibinfo {author} {\bibfnamefont {R.~E.}\ \bibnamefont
  {Prange}}\ and\ \bibinfo {author} {\bibfnamefont {T.-W.~W.}\ \bibnamefont
  {Nee}},\ }\href {\doibase 10.1103/PhysRev.168.779} {\bibfield  {journal}
  {\bibinfo  {journal} {Phys. Rev.}\ }\textbf {\bibinfo {volume} {168}},\
  \bibinfo {pages} {779} (\bibinfo {year} {1968})}\BibitemShut {NoStop}%
\bibitem [{\citenamefont {Te{\v{s}}anovi{\'{c}}}\ \emph
  {et~al.}(1986)\citenamefont {Te{\v{s}}anovi{\'{c}}}, \citenamefont
  {Jari{\'{c}}},\ and\ \citenamefont {Maekawa}}]{Tesanovic1986}%
  \BibitemOpen
  \bibfield  {author} {\bibinfo {author} {\bibfnamefont {Z.}~\bibnamefont
  {Te{\v{s}}anovi{\'{c}}}}, \bibinfo {author} {\bibfnamefont {M.~V.}\
  \bibnamefont {Jari{\'{c}}}}, \ and\ \bibinfo {author} {\bibfnamefont
  {S.}~\bibnamefont {Maekawa}},\ }\href {\doibase 10.1103/PhysRevLett.57.2760}
  {\bibfield  {journal} {\bibinfo  {journal} {Phys. Rev. Lett.}\ }\textbf
  {\bibinfo {volume} {57}},\ \bibinfo {pages} {2760} (\bibinfo {year}
  {1986})}\BibitemShut {NoStop}%
\bibitem [{\citenamefont {Trivedi}\ and\ \citenamefont
  {Ashcroft}(1988)}]{Trivedi1988}%
  \BibitemOpen
  \bibfield  {author} {\bibinfo {author} {\bibfnamefont {N.}~\bibnamefont
  {Trivedi}}\ and\ \bibinfo {author} {\bibfnamefont {N.~W.}\ \bibnamefont
  {Ashcroft}},\ }\href {\doibase 10.1103/PhysRevB.38.12298} {\bibfield
  {journal} {\bibinfo  {journal} {Phys. Rev. B}\ }\textbf {\bibinfo {volume}
  {38}},\ \bibinfo {pages} {12298} (\bibinfo {year} {1988})}\BibitemShut
  {NoStop}%
\bibitem [{\citenamefont {Meyerovich}\ and\ \citenamefont
  {Stepaniants}(1994)}]{Meyerovich1994}%
  \BibitemOpen
  \bibfield  {author} {\bibinfo {author} {\bibfnamefont {A.~E.}\ \bibnamefont
  {Meyerovich}}\ and\ \bibinfo {author} {\bibfnamefont {S.}~\bibnamefont
  {Stepaniants}},\ }\href {\doibase 10.1103/PhysRevLett.73.316} {\bibfield
  {journal} {\bibinfo  {journal} {Phys. Rev. Lett.}\ }\textbf {\bibinfo
  {volume} {73}},\ \bibinfo {pages} {316} (\bibinfo {year} {1994})}\BibitemShut
  {NoStop}%
\bibitem [{\citenamefont {Jacoboni}(2010)}]{Jacoboni2010}%
  \BibitemOpen
  \bibfield  {author} {\bibinfo {author} {\bibfnamefont {C.}~\bibnamefont
  {Jacoboni}},\ }\href {\doibase 10.1007/978-3-642-10586-9} {\emph {\bibinfo
  {title} {{Theory of Electron Transport in Semiconductors: A Pathway from
  Elementary Physics to Nonequilibrium Green Functions}}}},\ Vol.\ \bibinfo
  {volume} {165}\ (\bibinfo  {publisher} {Springer Science \& Business Media},\
  \bibinfo {year} {2010})\BibitemShut {NoStop}%
\bibitem [{\citenamefont {Moors}\ \emph {et~al.}(2014)\citenamefont {Moors},
  \citenamefont {Sor{\'{e}}e}, \citenamefont {Tőkei},\ and\ \citenamefont
  {Magnus}}]{Moors2014}%
  \BibitemOpen
  \bibfield  {author} {\bibinfo {author} {\bibfnamefont {K.}~\bibnamefont
  {Moors}}, \bibinfo {author} {\bibfnamefont {B.}~\bibnamefont {Sor{\'{e}}e}},
  \bibinfo {author} {\bibfnamefont {Z.}~\bibnamefont {Tőkei}}, \ and\ \bibinfo
  {author} {\bibfnamefont {W.}~\bibnamefont {Magnus}},\ }\href {\doibase
  10.1063/1.4892984} {\bibfield  {journal} {\bibinfo  {journal} {J. Appl.
  Phys.}\ }\textbf {\bibinfo {volume} {116}},\ \bibinfo {pages} {063714}
  (\bibinfo {year} {2014})}\BibitemShut {NoStop}%
\bibitem [{\citenamefont {Moors}\ \emph {et~al.}(2016)\citenamefont {Moors},
  \citenamefont {Sor{\'{e}}e},\ and\ \citenamefont {Magnus}}]{Moors2016}%
  \BibitemOpen
  \bibfield  {author} {\bibinfo {author} {\bibfnamefont {K.}~\bibnamefont
  {Moors}}, \bibinfo {author} {\bibfnamefont {B.}~\bibnamefont {Sor{\'{e}}e}},
  \ and\ \bibinfo {author} {\bibfnamefont {W.}~\bibnamefont {Magnus}},\ }\href
  {\doibase 10.1088/0953-8984/28/36/365302} {\bibfield  {journal} {\bibinfo
  {journal} {J. Phys.: Condens. Matter}\ }\textbf {\bibinfo {volume} {28}},\
  \bibinfo {pages} {365302} (\bibinfo {year} {2016})}\BibitemShut {NoStop}%
\bibitem [{\citenamefont {Goharrizi}\ \emph {et~al.}(2011)\citenamefont
  {Goharrizi}, \citenamefont {Pourfath}, \citenamefont {Fathipour},
  \citenamefont {Kosina},\ and\ \citenamefont {Selberherr}}]{Goharrizi2011}%
  \BibitemOpen
  \bibfield  {author} {\bibinfo {author} {\bibfnamefont {A.~Y.}\ \bibnamefont
  {Goharrizi}}, \bibinfo {author} {\bibfnamefont {M.}~\bibnamefont {Pourfath}},
  \bibinfo {author} {\bibfnamefont {M.}~\bibnamefont {Fathipour}}, \bibinfo
  {author} {\bibfnamefont {H.}~\bibnamefont {Kosina}}, \ and\ \bibinfo {author}
  {\bibfnamefont {S.}~\bibnamefont {Selberherr}},\ }\href {\doibase
  10.1109/TED.2011.2163719} {\bibfield  {journal} {\bibinfo  {journal} {IEEE
  Trans. Electron Devices}\ }\textbf {\bibinfo {volume} {58}},\ \bibinfo
  {pages} {3725} (\bibinfo {year} {2011})}\BibitemShut {NoStop}%
\bibitem [{\citenamefont {Hwang}\ and\ \citenamefont {{Das
  Sarma}}(2008)}]{Hwang2008B}%
  \BibitemOpen
  \bibfield  {author} {\bibinfo {author} {\bibfnamefont {E.~H.}\ \bibnamefont
  {Hwang}}\ and\ \bibinfo {author} {\bibfnamefont {S.}~\bibnamefont {{Das
  Sarma}}},\ }\href {\doibase 10.1103/PhysRevB.77.115449} {\bibfield  {journal}
  {\bibinfo  {journal} {Phys. Rev. B}\ }\textbf {\bibinfo {volume} {77}},\
  \bibinfo {pages} {115449} (\bibinfo {year} {2008})}\BibitemShut {NoStop}%
\bibitem [{\citenamefont {{Das Sarma}}\ \emph {et~al.}(2011)\citenamefont {{Das
  Sarma}}, \citenamefont {Adam}, \citenamefont {Hwang},\ and\ \citenamefont
  {Rossi}}]{DasSarma2011}%
  \BibitemOpen
  \bibfield  {author} {\bibinfo {author} {\bibfnamefont {S.}~\bibnamefont {{Das
  Sarma}}}, \bibinfo {author} {\bibfnamefont {S.}~\bibnamefont {Adam}},
  \bibinfo {author} {\bibfnamefont {E.~H.}\ \bibnamefont {Hwang}}, \ and\
  \bibinfo {author} {\bibnamefont {Rossi}},\ }\href {\doibase
  10.1103/RevModPhys.83.407} {\bibfield  {journal} {\bibinfo  {journal} {Rev.
  Mod. Phys.}\ }\textbf {\bibinfo {volume} {83}},\ \bibinfo {pages} {407}
  (\bibinfo {year} {2011})}\BibitemShut {NoStop}%
\bibitem [{\citenamefont {Reich}\ \emph {et~al.}(2002)\citenamefont {Reich},
  \citenamefont {Maultzsch}, \citenamefont {Thomsen},\ and\ \citenamefont
  {Ordejon}}]{Reich2002}%
  \BibitemOpen
  \bibfield  {author} {\bibinfo {author} {\bibfnamefont {S.}~\bibnamefont
  {Reich}}, \bibinfo {author} {\bibfnamefont {J.}~\bibnamefont {Maultzsch}},
  \bibinfo {author} {\bibfnamefont {C.}~\bibnamefont {Thomsen}}, \ and\
  \bibinfo {author} {\bibfnamefont {P.}~\bibnamefont {Ordejon}},\ }\href
  {\doibase 10.1103/PhysRevB.66.035412} {\bibfield  {journal} {\bibinfo
  {journal} {Phys. Rev. B}\ }\textbf {\bibinfo {volume} {66}},\ \bibinfo
  {pages} {035412} (\bibinfo {year} {2002})}\BibitemShut {NoStop}%
\bibitem [{Note1()}]{Note1}%
  \BibitemOpen
  \bibinfo {note} {We dismiss the nonphysical solution with $\kappa = 0$
  ($\kappa ^\prime = 0$).}\BibitemShut {Stop}%
\end{thebibliography}%

\appendix

\section{Simplified models}
\label{appendix:simplified_models}
\subsection{Nanoribbons} \label{subsec:GNR_simplified}
Brey and Fertig introduced the following 2D linear four-band (Dirac-like) model to model the surface and edge states of zigzag and armchair GNRs:\cite{Brey2006B}
\begin{align}
	\hat{H} = \hbar \vF \lef \begin{matrix}
		0 & k_x + \imu k_z & 0 & 0 \\
		k_x - \imu k_z & 0 & 0 & 0 \\
		0 & 0 & 0 & -k'_x + \imu k'_z \\
		0 & 0 & -k'_x - \imu k'_z & 0 \\
	\end{matrix} \rig ,
\end{align}
approximating the spectrum of the conventional nearest-neighbor tight-binding model for graphene\cite{Reich2002} near the Dirac points of the $K$ and $K'$ valleys. The wave vectors $\veck$ and $\veck^\prime$ denote the separation in reciprocal space from the $\vecK$ and $\vecK'$ points, respectively.
The solutions of this Hamiltonian can be written as envelope functions organized in a four-vector $\Psi = (\psi_A, \psi_B, -\psi'_A, - \psi'_B)$ ($A$ and $B$ denoting the two sublattices), having energy $E$ and normalized on each sublattice separately:\cite{Ryu2002,Brey2006A}
\begin{align}
	\int \deriv^2 r \; \left[ |\psi_\mu(\vecr)|^2 + |\psi'_\mu(\vecr)|^2 \right] = 1/2, \quad (\mu = A, B).
\end{align}
For a GNR, the wave function should be confined to the ribbon geometry, which translates into specific boundary conditions for the wave function solutions. The general confined solutions can be written in the form:
\begin{align} \label{eq:sol_general}
	\psi_\mu(\vecr) \equiv \e^{\imu k_\parallel r_\parallel} \, \phi_\mu(r_\perp),
		\quad \psi_\mu(\vecr) \equiv \e^{\imu k_\parallel^\prime r_\parallel} \, \phi_\mu^\prime(r_\perp),
\end{align}
where $r_\parallel$ ($r_\perp$) denotes the direction along (perpendicular to) the transport direction in the plane of the ribbon.
For the transverse part of the wave function, we have the following general form:
\begin{align}
	\phi_\mu (r_\perp) = C_\mu \, \e^{\kappa r_\perp} + D_\mu \, \e^{- \kappa r_\perp}, \quad
		\tilde{E}^2 = k_\parallel^2 - \kappa^2,
\end{align}
with $C_\mu$, $D_\mu$, and $\kappa$ in general complex and $\tilde{E} \equiv E/(\hbar \vF)$. Note that exactly the same form can be considered for $\Psi^\prime(\vecr)$, with $C_\mu^\prime$, $D_\mu^\prime$, and $\kappa^\prime$. At the extremities of the opposite edges of a zigzag ribbon, the carbon atoms belong to the opposite sublattice type. Confinement within a zigzag ribbon thus requires the wave function of opposite sublattice type ($\mu = A, B$) to vanish on the opposite edges, leading to the following transcendental equations for $\kappa$ and $\kappa^\prime$:
\begin{align}
	\frac{k_\parallel - \kappa}{k_\parallel + \kappa} = \e^{- 2 W \kappa}, \qquad
			\frac{k_\parallel^\prime + \kappa^\prime}{k_\parallel^\prime - \kappa^\prime} = \e^{- 2 W \kappa^\prime}.
\end{align}
When $k_\parallel > 1/W$ ($k_\parallel^\prime < -1/W$), there are solutions\footnote{We dismiss the nonphysical solution with $\kappa = 0$ ($\kappa^\prime = 0$).} for $\kappa$ ($\kappa^\prime$) being real, redefined as $\kappa \equiv 1/w$, with $\tilde{E}(k_\parallel, w) = \pm \sqrt{k_\parallel^2 - 1/w^2}$ and $|w|$ a measure of the edge state width. When $1/w \gg 1/W$ ($1/w^\prime \ll -1/W$), we obtain $1/w \rightarrow k_\parallel$ ($1/w^\prime \rightarrow -k_\parallel^\prime$) and $\tilde{E}(k_\parallel, w) \rightarrow 0$ [$\tilde{E}(k_\parallel^\prime, w^\prime) \rightarrow 0$]. These (approximately) zero-energy solutions for the $K$ and $K^\prime$ valleys are the equivalent of the nearly flat edge-state bands that are retrieved with the pseudopotential method and connect both valleys. This connection between the two valleys cannot be retrieved with the simplified model presented here, because the valleys are considered to be uncoupled. We therefore glue the simplified edge-state solutions together halfway in between $\vecK$ and $\vecK^\prime$ along $k_\parallel$ in an \textit{ad hoc} manner.
There are also solutions for imaginary $\kappa$ ($\kappa^\prime$), redefined as $\kappa \equiv \imu k_n$ ($\kappa^\prime \equiv \imu k_n^\prime$), with $\tilde{E}(k_\parallel, k_n) = \pm \sqrt{k_\parallel^2 + k_n^2}$, for which the transcendental equations can be rewritten as:
\begin{align} \label{eq:zigzag_imaginary}
	k_\parallel = \frac{k_n}{\tan(k_n W)}, \qquad -k_\parallel^\prime = \frac{k_n^\prime}{\tan(k_n^\prime W)}.
\end{align}
The lowest energy solution of Eq.~\eqref{eq:zigzag_imaginary}, appearing when $k_n < 1/W$ ($k_n^\prime > - 1/W$), vanishes when $k_n$ ($k_n^\prime$) reaches $1/W$ ($-1/W$), transitioning into the solution for real $\kappa$ ($\kappa^\prime$). This is a transition from a ribbon bulk state to an edge state for a state with energy equal to $\pm 1/W$.
For an armchair ribbon, confinement implies that the wave function should vanish on both sublattices separately, as the edges consists of both $A$- and $B$-type carbon atoms. This can only be realized by mixing the two valleys, leading to the following standing wave solutions:
\begin{align}
\begin{split}
	\phi_\mu(r_\perp) &= \e^{\imu k_n r_\perp}, \qquad \phi_\mu^\prime(r_\perp) = \e^{- \imu k_n r_\perp}, \\
	\e^{\imu 2 k_n W} &= \e^{\imu \Delta K W},
\end{split}
\end{align}
where $\Delta K \equiv 2 \sqrt{3} \pi / (9 a_0)$, the minimal distance between $\vecK$ and $\vecK^\prime$ projected on the confinement direction ($k_\perp^\prime = k_\perp + \Delta K$).
The total width of an armchair ribbon is related to the total number of carbon atoms in the supercell $N$ (containing a single armchair piece along the transport direction) by $W = N \sqrt{3} a_0 / 4$. This implies:
\begin{align}
	2 k_n W = N \pi/6 + n 2 \pi = j 2 \pi/3 + n 2 \pi,
\end{align}
with integer $n$ and $j = -1, 0, 1$, determined by the relation:
\begin{align}
	N/4 = 3 M + j,
\end{align}
$M$ being an integer. The allowed values for the wave vector $k_n$ are thus given by:
\begin{align} \label{eq:armchair_subbands}
	k_n = j \frac{\pi}{3 W} + n \frac{\pi}{W} = \frac{4}{N} \frac{\pi}{\sqrt{3} a_0} (n + j/3),
\end{align}
with corresponding energies $\tilde{E}(k, k_n) = \pm \sqrt{k^2 + k_n^2}$. The armchair ribbon has a gapless spectrum when $j = 0$ and is insulating when $j = \pm 1$.

Note that this analysis is only for armchair ribbons with a uniform width along the transport direction, with $N/4$ being integer valued. One can also consider armchair ribbons with oscillating widths for which $N/4$ is not integer valued. Their solutions should equally satisfy the mixed boundary conditions, as the condition which governs the applicability of this type of boundary conditions is met by the amount of dangling bonds of $A$ and $B$ type of the edges being equal.\cite{Akhmerov2008} It is straightforward to verify that this is the case for any type of edge configuration as long as the supercell of the ribbon is aligned with an armchair direction.

The solutions of a randomly oriented ribbon (zigzag nor armchair) follow the same boundary conditions as those of a zigzag ribbon, their supercell not aligning with any armchair direction.\cite{Akhmerov2008} The $K$ and $K^\prime$ valleys do not overlap along the transport direction and there is no valley mixing. The only difference is the minimal difference between the $K$ and $K^\prime$ valleys projected along the transport direction. This difference is given by $\Delta K (\angleRibbon) = \Delta K \, \sin\angleRibbon$ [see Fig.~\subref*{fig:graphene_c}], with $\angleRibbon$ the angle of the GNR transport direction with respect to the armchair orientation.

The simplified model is able to reproduce the important features of the band structure of the different GNRs simulated with the pseudopotential method [compare Fig.~\subref*{fig:graphene_d} and Fig.~\ref{fig:GNR_Bands} for example]. There are notable differences with our pseudopotential or other atomistic approaches, such as the precise band gap of the armchair GNRs due to their claromatic behavior,\cite{Balaban2009,Fischetti2013} the small but finite dispersion of the edge state band, and the overall asymmetry between conduction and valence subbands, but these differences are minor and can safely be neglected for an analysis of the transport properties of the GNRs with widths and (large) doping levels under consideration here (as required for low-resistivity nanoscaled applications).

\subsection{Atomistic edge roughness} \label{subsec:AER_simplified}
To evaluate the resistivity due to AER scattering as a function of the ribbon width with the simplified GNR description of Appendix~\ref{subsec:GNR_simplified}, we propose the following simplified averages for the matrix elements squared of Eq.~\eqref{eq:ER_Matrix_El}:
\begin{align} \label{eq:AER_scat_simplified}
	\left\langle \left| \langle i \mid \VAER \mid f \rangle \right|^2 \right\rangle_\AER \! \! \! \! \!
			\rightarrow E_\AER^2 \frac{\AERSD^2}{W_i W_f} C_\AERCL(\Delta \veck_\parallel),
\end{align}
with $W_{i \, (f)}$ the width of the initial (final) state and $\AERSD$ the AER standard deviation. We consider $W_i, W_f$ to be equal to the GNR width $W$ for a bulk state and equal to the width $w$ for an edge state, as obtained from the transcendental equations in Appendix~\ref{subsec:GNR_simplified}. The details of the matrix elements in Eq.~\eqref{eq:ER_Matrix_El} that depend on the pseudopotential wave functions and edge roughness potentials have been replaced by a single energy parameter $E_\AER$, which can be fitted to retrieve the pseudopotential-based results. The dependency on the width of initial and final state and on the AER standard deviation are inspired by Ando's surface roughness model.\cite{Ando1982}

\end{document}